\documentclass[a4paper,11pt]{article}
\usepackage{jcappub} % for details on the use of the package, please see the JCAP-author-manual

\title{Dark energy and Equivalence Principle constraints from astrophysical tests of the stability of the fine-structure constant}

\author[a,b,1]{C. J. A. P. Martins,\note{Corresponding author.}}
\author[a,c]{A. M. M. Pinho,}
\author[a,c]{R. F. C. Alves,}
\author[d]{M. Pino,}
\author[e]{C. I. S. A. Rocha}
\author[f]{and M. von Wietersheim}

\affiliation[a]{Centro de Astrof\'{\i}sica da Universidade do Porto, Rua das Estrelas, 4150-762 Porto, Portugal}
\affiliation[b]{Instituto de Astrof\'{\i}sica e Ci\^encias do Espa\c co, CAUP, Rua das Estrelas, 4150-762 Porto, Portugal}
\affiliation[c]{Faculdade de Ci\^encias, Universidade do Porto, Rua do Campo Alegre 687, 4169-007 Porto, Portugal}
\affiliation[d]{Institut Dom\`enech i Montaner, C/Maspujols 21-23, 43206 Reus, Spain}
\affiliation[e]{Externato Ribadouro, Rua de Santa Catarina 1346, 4000-447 Porto, Portugal}
\affiliation[f]{Institut Manuel Sales i Ferr\'e, Avinguda de les Escoles 6, 43550 Ulldecona, Spain}

\emailAdd{Carlos.Martins@astro.up.pt}
\emailAdd{Ana.Pinho@astro.up.pt}
\emailAdd{up201106579@fc.up.pt}
\emailAdd{mpc\_97@yahoo.com}
\emailAdd{cisar97@hotmail.com}
\emailAdd{maxivonw@gmail.com}

\abstract{Astrophysical tests of the stability of fundamental couplings, such as the fine-structure constant $\alpha$, are becoming an increasingly powerful probe of new physics. Here we discuss how these measurements, combined with local atomic clock tests and Type Ia supernova and Hubble parameter data, constrain the simplest class of dynamical dark energy models where the same degree of freedom is assumed to provide both the dark energy and (through a dimensionless coupling, $\zeta$, to the electromagnetic sector) the $\alpha$ variation. Specifically, current data tightly constrains a combination of $\zeta$ and the present dark energy equation of state $w_0$. Moreover, in these models the new degree of freedom inevitably couples to nucleons (through the $\alpha$ dependence of their masses) and leads to violations of the Weak Equivalence Principle. We obtain indirect bounds on the E\"{o}tv\"{o}s parameter $\eta$ that are typically stronger than the current direct ones. We discuss the model-dependence of our results and briefly comment on how the forthcoming generation of high-resolution ultra-stable spectrographs will enable significantly tighter constraints.}

\keywords{Observational cosmology; Fine-structure constant; Dark energy; Equivalence principle}

\begin{document}
\maketitle
\flushbottom

%%%%%%%%%%%%%%%%%%%%%%%%%%%%%%%%%%%%%%%%%%%%%%%%%%%%%%%%%%%%%%%%%%%%%%%%%%%%%%%%%%
\section{Introduction}
\label{sec:intro}

The discovery of cosmic acceleration at the end of the last century, from luminosity distance measurements of Type Ia supernovas \cite{SN1,SN2}, has led to wide-ranging efforts trying to characterize its observational properties and understand its theoretical origin. While a cosmological constant $\Lambda$ remains the simplest explanation consistent with current observational data, the well-known fine-tuning problems associated with this solution imply that alternative scenarios should be sought and actively tested. The most natural alternative explanation would involve scalar fields, an example of which is the recently discovered Higgs field \cite{ATLAS,CMS}. Such cosmological scalar fields would lead to dynamical dark energy scenarios.

If dynamical scalar fields are indeed present, one naturally expects them to couple to the rest of the model, unless a yet-unknown symmetry is postulated to suppress these couplings \cite{Carroll}. In particular, a coupling of the field to the electromagnetic sector will lead to spacetime variations of the fine-structure constant $\alpha$---see \cite{uzanLR,cjmGRG} for recent reviews on this topic. Indeed there have been some recent indications of such a variation \cite{Webb}, at the relative level of variation of a few parts per million, and an ongoing dedicated Large Program at ESO's Very Large Telescope (VLT) is aiming to test them \cite{LP1,LP3}. Regardless of the outcome of these studies (ie, whether they provide detections of variations or just null results) the measurements have cosmological implications that go beyond the mere fundamental nature of the tests themselves. The goal of this work is precisely to address some of these cosmological implications.

Specifically, in the same spirit of \cite{Erminia1,Erminia2}, we discuss how astrophysical and local tests of the stability of $\alpha$ can be used as additional tests of the underlying dynamical dark energy scenarios. We will thus be making the minimal assumption that the same dynamical degree of freedom is responsible for the dark energy and the $\alpha$ variations---there are known as Class I models in the classification of \cite{cjmGRG}. In this case any observational (astrophysical) or experimental (local) tests of the stability of $\alpha$ will directly constrain dark energy. The future impact of these methods as a dark energy probe has recently been assessed in some detail \cite{Amendola,Leite,LeiteNEW}, in preparation for forthcoming facilities which include these measurements as a key science driver (we will return to these at the end of the paper). Our goal here is to describe how current data already yields useful constraints on Class I models, and thus provide a proof of concept for these future facilities. (Any model not in Class I is called a Class II model, and we will also briefly comment on them later.)

In a previous work \cite{Pinho} we obtained a non-trivial constraint on the dimensionless coupling of the scalar field to the electromagnetic sector, $\zeta$, at the two-sigma ($95.4\%$) confidence level (and marginalizing over the dark energy equation of state $w_0$)
\begin{equation} \label{zetaboundfix}
|\zeta|<5\times10^{-6}\,,
\end{equation}
significantly improving upon previous constraints, while the 1D likelihood for $w_0$ (marginalizing over $\zeta$) was found to be, at the three-sigma ($99.7\%$) confidence level
\begin{equation} \label{w0boundfix}
-1.05<w_0<-0.94\,.
\end{equation}
Both of these were obtained by assuming a fiducial model for the dark energy with a constant equation of state, $w(z)=w_0$. Here we will relax this assumption, and study models where the dark energy equation of state does vary with redshift. Presumably these will be somewhat more realistic, but we have also aimed to preserve conceptual simplicity by choosing parametrizations that do not increase the number of free parameters. Studying and quantifying how the above constraints depend on the choice of models (and priors) is one of the goals of the present work.

Our second goal is to show that, since in these models the new degree of freedom inevitably couples to nucleons (through the $\alpha$ dependence of their masses) and thereby leads to violations of the Weak Equivalence Principle \cite{Dvali,Chiba}, the new constraints on $\zeta$ also correspond to very tight constraints on the E\"{o}tv\"{o}s parameter $\eta$ quantifying violations of the Weak Equivalence Principle (WEP). Although these constraints are indirect and model-dependent (since Class I models are assumed) they are comparable to or even tighter than current direct constraints. Thus forthcoming tests of the WEP will also provide important consistency tests of these models.

In what follows we start by briefly reviewing the relation between a varying $\alpha$ and dynamical energy in the case of Class I models, as well as the relation to the WEP violations. We then list the local, astrophysical and cosmological datasets we will be considering in our analysis, and proceed to study two representative classes of models, constraining them with the aforementioned data. In particular we discuss how the obtained constraints depend on the model and the choice of priors. Finally we present some conclusions, as well as a brief discussion of the improvements on the sensitivity of these tests expected from future facilities. We also provide a brief discussion, in an appendix, of the usage of the Oklo bound to further constrain these models.

%%%%%%%%%%%%%%%%%%%%%%%%%%%%%%%%%%%%%%%%%%%%%%%%%%%%%%%%%%%%%%%%%%%%%%%%%%%%%%%%%%
\section{Varying $\alpha$, dark energy and the Weak Equivalence Principle}
\label{sec:coupling}

Dynamical scalar fields in an effective 4D field theory are naturally expected to couple to the rest of the theory, unless a (still unknown) symmetry is postulated to suppress this coupling \cite{Carroll,Dvali,Chiba}. In what follows we will assume this to be the case for the dynamical degree of freedom responsible for the dark energy. Specifically we will assume a coupling between the scalar field, denoted $\phi$, and the electromagnetic sector, which stems from a gauge kinetic function $B_F(\phi)$
\begin{equation}
{\cal L}_{\phi F} = - \frac{1}{4} B_F(\phi) F_{\mu\nu}F^{\mu\nu}\,.
\end{equation}
One can assume this function to be linear,
\begin{equation}
B_F(\phi) = 1- \zeta \kappa (\phi-\phi_0)\,,
\end{equation}
(with $\kappa^2=8\pi G$) since, as has been pointed out in \cite{Dvali}, the absence of such a term would require the presence a $\phi\to-\phi$ symmetry, but such a symmetry must be broken throughout most of the cosmological evolution. As is physically clear, the relevant parameter in the cosmological evolution is the field displacement relative to its present-day value (in particular $\phi_0$ could be set to zero). In these models the proton and neutron masses are also expected to vary, due to the electromagnetic corrections of their masses; while we will not discuss this in detail in the present work, one relevant consequence of this fact is that local tests of the Equivalence Principle lead to the conservative constraint on the dimensionless coupling parameter (see \cite{uzanLR} for an overview)
\begin{equation}
|\zeta_{\rm local}|<10^{-3}\,,\label{localzeta}
\end{equation}
while in \cite{Erminia1} an independent few-percent constraint on this coupling was obtained using CMB and large-scale structure data in combination with direct measurements of the expansion rate of the universe.

We note that there is in principle an additional source term driving the evolution of the scalar field, due to a $F^2B_F'$ term. By comparison to the standard (kinetic and potential energy) terms, the contribution of this term is expected to be subdominant, both because its average is zero for a radiation fluid and because the corresponding term for the baryonic density is constrained by the same reasons discussed in the previous paragraph. For these reasons, in what follows we neglect this term, which would lead to spatial/environmental dependencies. We nevertheless note that this term can play a role in scenarios where the dominant standard term is suppressed.

With these assumptions one can explicitly relate the evolution of $\alpha$ to that of dark energy, as in \cite{Erminia1} whose derivation we summarize here. The evolution of $\alpha$ can be written
\begin{equation}
\frac{\Delta \alpha}{\alpha} \equiv \frac{\alpha-\alpha_0}{\alpha_0} =B_F^{-1}(\phi)-1=
\zeta \kappa (\phi-\phi_0) \,,
\end{equation}
and defining the fraction of the dark energy density
\begin{equation}
\Omega_\phi (z) \equiv \frac{\rho_\phi(z)}{\rho_{\rm tot}(z)} \simeq \frac{\rho_\phi(z)}{\rho_\phi(z)+\rho_m(z)} \,,
\end{equation}
where in the last step we have neglected the contribution from radiation (since we will be interested in low redshifts, $z<5$, where it is indeed negligible), the evolution of the putative scalar field can be expressed in terms of the dark energy properties $\Omega_\phi$ and $w_\phi$ as \cite{Nunes}
\begin{equation}
1+w_\phi = \frac{(\kappa\phi')^2}{3 \Omega_\phi} \,,
\end{equation}
with the prime denoting the derivative with respect to the logarithm of the scale factor. We finally obtain
\begin{equation} \label{eq:dalfa}
\frac{\Delta\alpha}{\alpha}(z) =\zeta \int_0^{z}\sqrt{3\Omega_\phi(z)\left(1+w_\phi(z)\right)}\frac{dz'}{1+z'}\,.
\end{equation}
The above relation assumes a canonical scalar field, but the argument can be repeated for phantom fields \cite{Phantom}, leading to 
\begin{equation} \label{eq:dalfa2}
\frac{\Delta\alpha}{\alpha}(z) =-\zeta \int_0^{z}\sqrt{3\Omega_\phi(z)\left|1+w_\phi(z)\right|}\frac{dz'}{1+z'}\,;
\end{equation}
the change of sign stems from the fact that one expects phantom filed to roll up the potential rather than down.

The realization that varying fundamental couplings induce violations of the universality of free fall goes back at least to the work of Dicke---we refer the reader to \cite{Damour} for a recent thorough discussion. In our present context, the key point is that a light scalar field such as we are considering inevitably couples to nucleons due to the $\alpha$ dependence of their masses, and therefore it mediates an isotope-dependent long-range force. This can be simply quantified through the dimensionless E\"{o}tv\"{o}s parameter $\eta$, which describes the level of violation of the Weak Equivalence Principle. One can show that for the class of models we are considering the  E\"{o}tv\"{o}s parameter and the dimensionless coupling $\zeta$ are simply related by \cite{Dvali,Chiba,Damour,uzanLR}
\begin{equation} \label{eq:eotvos}
\eta \approx 10^{-3}\zeta^2\,;
\end{equation}
therefore, the constraint on $\zeta$ obtained in \cite{Pinho} and reproduced in Eq. \ref{zetaboundfix} leads to the two-sigma indirect bound
\begin{equation} \label{etaboundfix}
\eta<2.5\times10^{-14}\,,
\end{equation}
which is stronger than the current direct bounds that will be discussed in the following section. We emphasize that this relation only applies to Class I models, and as we will see in later sections the numerical pre-factor in Eq. \ref{etaboundfix} will be slightly different for different models within this class.

Two interesting examples of Class II models for which the relation between the E\"{o}tv\"{o}s parameter and the model's coupling parameter is different have been described and studied in the literature. One is the runaway dilaton class of models of Damour {\it et al.} \cite{DPV}, for which up-to-date constraints were recently obtained in \cite{NewDPV}. Another one is provided by Bekenstein-type scenarios such as the so-called Bekenstein-Sandvik-Barrow-Magueijo (BSBM) model \cite{BSBM}. Constraints on $\zeta$ in this model were recently updated by \cite{Leal} who found, at the two-sigma level and from astrophysical measurements of $\alpha$ alone
\begin{equation} \label{zetaboundbsbm}
|\zeta_{\small BSBM}|<2.2\times10^{-5}\,;
\end{equation}
in this case, using the relation between $\eta$ and $\zeta$ originally provided by \cite{BSBM}, this constraint leads to
\begin{equation} \label{etaboundbsbm}
\eta_{\small BSBM}<5\times10^{-14}\,;
\end{equation}
we will briefly return to these models in what follows. In general one can say that in Class II models the proportionality factor between $\eta$ and the square of the relevant coupling is smaller that $10^{-3}$, and therefore the couplings in such models are less constrained by WEP tests than those of Class I models.

%%%%%%%%%%%%%%%%%%%%%%%%%%%%%%%%%%%%%%%%%%%%%%%%%%%%%%%%%%%%%%%%%%%%%%%%%%%%%%%%%%
\begin{figure}[tbp]
\centering
\includegraphics[width=3in]{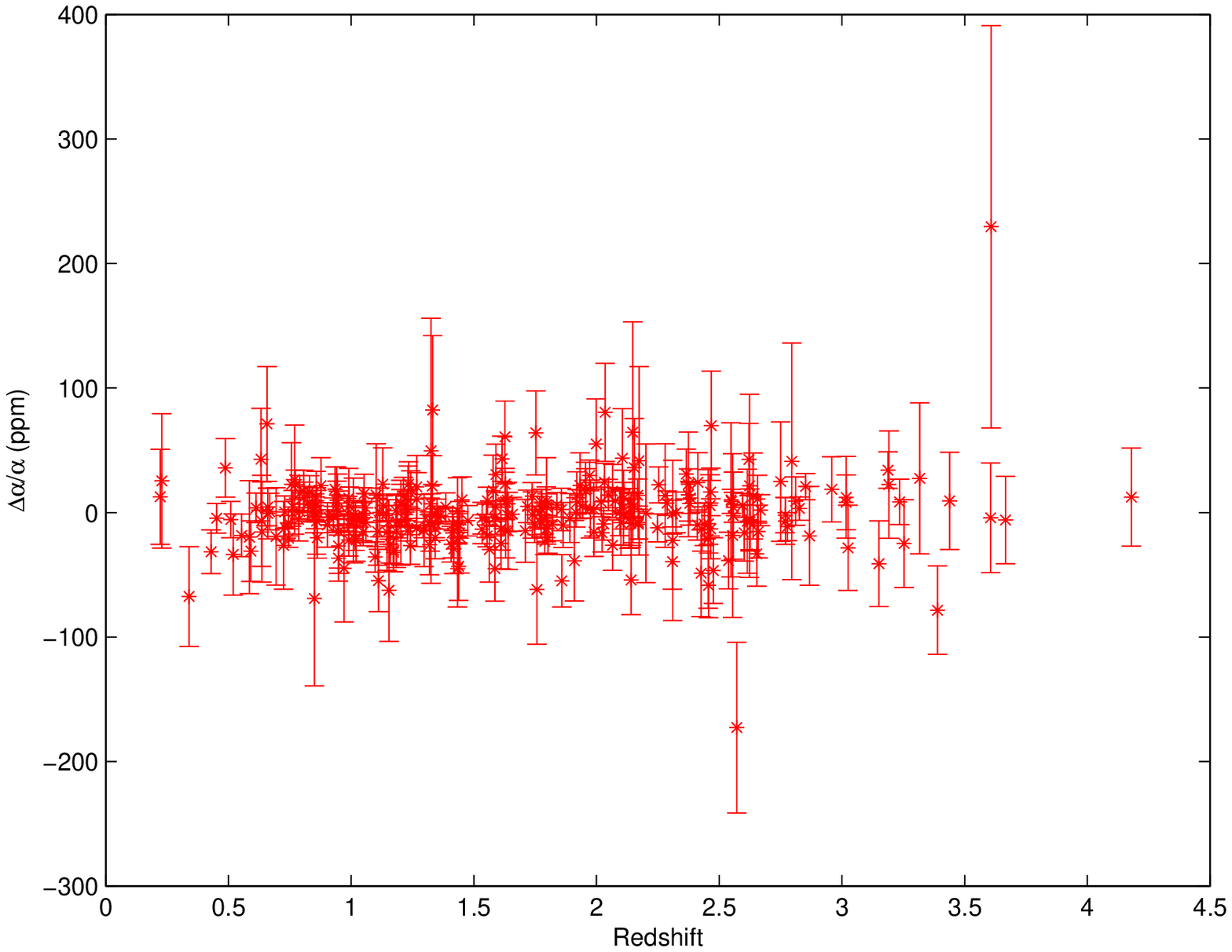}
\includegraphics[width=3in]{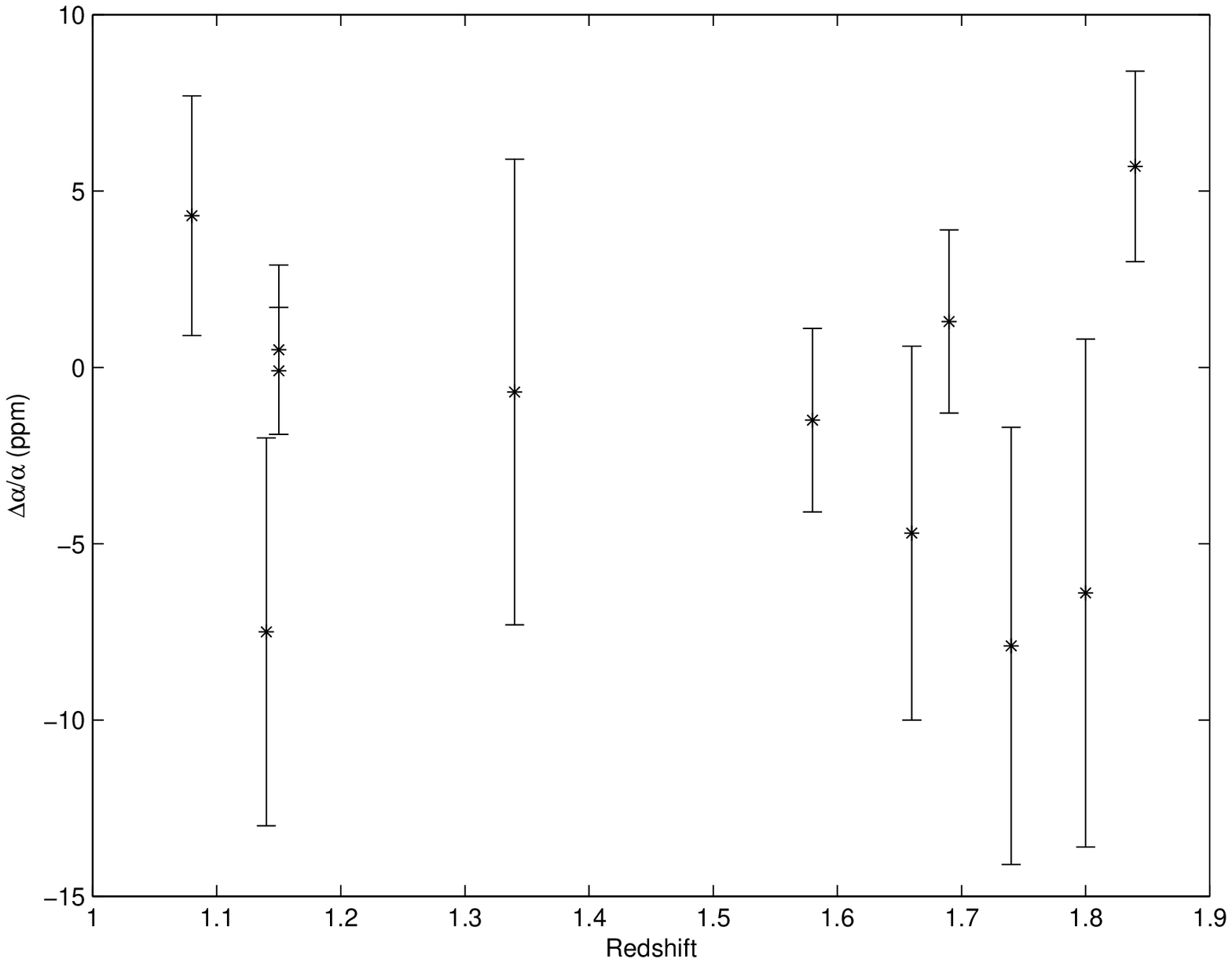}
\caption{\label{fig01}Currently available fine-structure constant measurements, with the relative values $\Delta\alpha/\alpha$ plotted as a function of redshift. The data of \protect\cite{Webb} is shown on the left panel, while the more recent data of Table \protect\ref{table1} is shown on the right panel. In both cases the error bars include both statistical and systematic uncertainties, added in quadrature. Note the difference in the vertical scales of both panels.}
\end{figure}
%%%%%%%%%%%%%%%%%%%%%%%%%%%%%%%%%%%%%%%%%%%%%%%%%%%%%%%%%%%%%%%%%%%%%%%%%%%%%%%%%%

%%%%%%%%%%%%%%%%%%%%%%%%%%%%%%%%%%%%%%%%%%%%%%%%%%%%%%%%%%%%%%%%%%%%%%%%%%%%%%%%%%
\section{Available datasets}
\label{sec:data}

Our goal is to constrain dynamical dark energy models coupled to the electromagnetic sector, by using the following datasets
\begin{itemize}
\item Cosmological data: we will use the Union2.1 dataset of 580 Type Ia supernovas \cite{Union} and the compilation of 28 Hubble parameter measurements from Farooq \& Ratra \cite{Farooq}. These datasets are insensitive to $\zeta$. (Strictly speaking, a varying $\alpha$ does affect the luminosity of Type Ia supernovas, but as recently shown in \cite{Erminia2} for parts-per-million level $\alpha$ variations the effect is too small to have an impact on current datasets, and we therefore neglect it in the present analysis.) However, this data does constrain the dark energy equation of state $w_0$, effectively providing us with a prior on it.
\item Laboratory data: we will use the atomic clock constraint on the current drift of $\alpha$ of Rosenband {\it et al.} \cite {Rosenband},
\begin{equation} \label{clocks0}
\frac{\dot\alpha}{\alpha} =(-1.6\pm2.3)\times10^{-17}\,{\rm yr}^{-1}\,.
\end{equation}
which we can also write in a dimensionless form by dividing by the present-day Hubble parameter,
\begin{equation} \label{clocks}
\frac{1}{H_0}\frac{\dot\alpha}{\alpha} =(-2.2\pm3.2)\times10^{-7}\,.
\end{equation}
This is the tightest available laboratory constraint on $\alpha$ only. Other laboratory constraints are weaker and also depend on other couplings too; we leave the discussion of these for subsequent work, but the interested reader can find overviews of atomic clock tests in \cite{Luo,Ferreira}

For the models under consideration this translates into
\begin{equation} \label{clocks2}
\frac{1}{H_0}\frac{\dot\alpha}{\alpha} =-\, \Sigma\, \zeta\sqrt{3\Omega_{\phi0}|1+w_0|}\,,
\end{equation}
where $\Sigma$ denotes the sign of $(1+w_0)$, so it is $+1$ for canonical fields and $-1$ for phantom fields.
\item Astrophysical data: we will use both the spectroscopic measurements of $\alpha$ of Webb {\it et al.} \cite{Webb} (a large dataset of 293 archival data measurements) and the smaller but more recent dataset of 11 dedicated measurements listed in Table \ref{table1}. The latter include the early results of the UVES Large Program for Testing Fundamental Physics \cite{LP1,LP3}, which is expected to be the one with a better control of possible systematics. Figure \ref{fig01} depicts these datasets.
\end{itemize}

%%%%%%%%%%%%%%%%%%%%%%%%%%%%%%%%%%%%%%%%%%%%%%%%%%%%%%%%%%%%%%%%%%%%%%%%%%%%%%%%%%
\begin{table}[tbp]
\centering
\begin{tabular}{|c|c|c|c|c|}
\hline
 Object & z & ${ \Delta\alpha}/{\alpha}$ (ppm) & Spectrograph & Ref. \\ 
\hline\hline
3 sources & 1.08 & $4.3\pm3.4$ & HIRES & \protect\cite{Songaila} \\
\hline
HS1549$+$1919 & 1.14 & $-7.5\pm5.5$ & UVES/HIRES/HDS & \protect\cite{LP3} \\
\hline
HE0515$-$4414 & 1.15 & $-0.1\pm1.8$ & UVES & \protect\cite{alphaMolaro} \\
\hline
HE0515$-$4414 & 1.15 & $0.5\pm2.4$ & HARPS/UVES & \protect\cite{alphaChand} \\
\hline
HS1549$+$1919 & 1.34 & $-0.7\pm6.6$ & UVES/HIRES/HDS & \protect\cite{LP3} \\
\hline
HE0001$-$2340 & 1.58 & $-1.5\pm2.6$ &  UVES & \protect\cite{alphaAgafonova}\\
\hline
HE1104$-$1805A & 1.66 & $-4.7\pm5.3$ & HIRES & \protect\cite{Songaila} \\
\hline
HE2217$-$2818 & 1.69 & $1.3\pm2.6$ &  UVES & \protect\cite{LP1}\\
\hline
HS1946$+$7658 & 1.74 & $-7.9\pm6.2$ & HIRES & \protect\cite{Songaila} \\
\hline
HS1549$+$1919 & 1.80 & $-6.4\pm7.2$ & UVES/HIRES/HDS & \protect\cite{LP3} \\
\hline
Q1101$-$264 & 1.84 & $5.7\pm2.7$ &  UVES & \protect\cite{alphaMolaro}\\
\hline
\end{tabular}
\caption{\label{table1}Recent dedicated measurements of $\alpha$. Listed are, respectively, the object along each line of sight, the redshift of the measurement, the measurement itself (in parts per million), the spectrograph, and the original reference. The first measurement is the weighted average from 8 absorbers in the redshift range $0.73<z<1.53$ along the lines of sight of HE1104-1805A, HS1700+6416 and HS1946+7658, reported in \cite{Songaila} without the values for individual systems. The UVES, HARPS, HIRES and HDS spectrographs are respectively in the VLT, ESO 3.6m, Keck and Subaru telescopes.}
\end{table}
%%%%%%%%%%%%%%%%%%%%%%%%%%%%%%%%%%%%%%%%%%%%%%%%%%%%%%%%%%%%%%%%%%%%%%%%%%%%%%%%%%

We use these datasets to constrain the dynamical dark energy models described in the following sections. The behavior of $\alpha$ will be determined by Eq.(\ref{eq:dalfa}) for canonical equations of state ($w(z)>-1$) and Eq.(\ref{eq:dalfa2}) for phantom equations of state ($w(z)<-1$). While in \cite{Pinho} we assumed a model with a constant equation of state $w(z)=w_0$, here we relax this assumption and study models where the equation of state is redshift-dependent, but nevertheless characterized by a single parameter $w_0$. This is done in the interest of conceptual simplicity and also because the currently available data can only weakly constrain models with additional free parameters. In any case, this does not prevent us from studying a fairly broad class of models.

Our main interest is in obtaining constraints on the $\zeta$--$w_0$ plane, and for this reason we will fix the Hubble parameter to be $H_0=70$ km/s/Mpc and the matter density to be $\Omega_{m0}=0.3$ (and further assume a flat universe, so $\Omega_{\phi0}=0.7$). This choice of cosmological parameters is fully consistent with the supernova and Hubble parameter data we use. Moreover, we have explicitly verified that allowing $H_0$, $\Omega_m$ or the curvature parameter to vary (within observationally reasonable ranges) and marginalizing over them does not significantly change our results. (This should be clear from the fact that a parts-per-million variation of $\alpha$ cannot dramatically affect these cosmological parameters.) It is clear that the critical cosmological parameter here is $w_0$ itself, as in Class I models it will be correlated with $\zeta$. We therefore consider 2D grids of $\zeta$ and $w_0$ values, and use standard maximum likelihood techniques to compare the models and the data. 

We also list here the available direct constraints on the dimensionless E\"{o}tv\"{o}s parameter, quantifying violations to the Weak Equivalence Principle. These stem from torsion balance tests, leading to \cite{Torsion}
\begin{equation}\label{boundetaT}
\eta=(-0.7\pm1.3)\times10^{-13}\,,
\end{equation}
while from lunar laser ranging one finds \cite{Lunar}
\begin{equation}\label{boundetaL}
\eta=(-0.8\pm1.2)\times10^{-13}\,.
\end{equation}
Both of these are quoted with their one-sigma uncertainties.

%%%%%%%%%%%%%%%%%%%%%%%%%%%%%%%%%%%%%%%%%%%%%%%%%%%%%%%%%%%%%%%%%%%%%%%%%%%%%%%%%%
\section{The thawing models of Slepian {\it et al.}}
\label{sec:sgz}

We will start by studying a dark energy parametrization which is relatively simple, in the sense that it does not involve any more parameters than we already have. The model was recently introduced by Slepian {\it et al.} \cite{SGZ}. The Friedmann equation has the following form
\begin{equation}\label{newsqrt}
\frac{H^2(z)}{H^2_0}=\Omega_{m}(1+z)^{3}+\Omega_{\phi}\left[\frac{(1+z)^{3}}{\Omega_{m}(1+z)^{3}+\Omega_{\phi}}\right]^{\frac{1+ w_0}{\Omega_{\phi}}} \,.
\end{equation}
We will be assuming flat universes, so $\Omega_m+\Omega_\phi=1$, and the model is therefore characterized by three independent parameters: $H_0$, $\Omega_m$ (which here we keep fixed as previously justified) and $w_0$. Note that the physical interpretation of these parameters is exactly the standard one, and in particular, $w_0$ is still the value of the dark energy equation of state {\it today}.

\begin{figure}[tbp]
\centering
\includegraphics[width=3in]{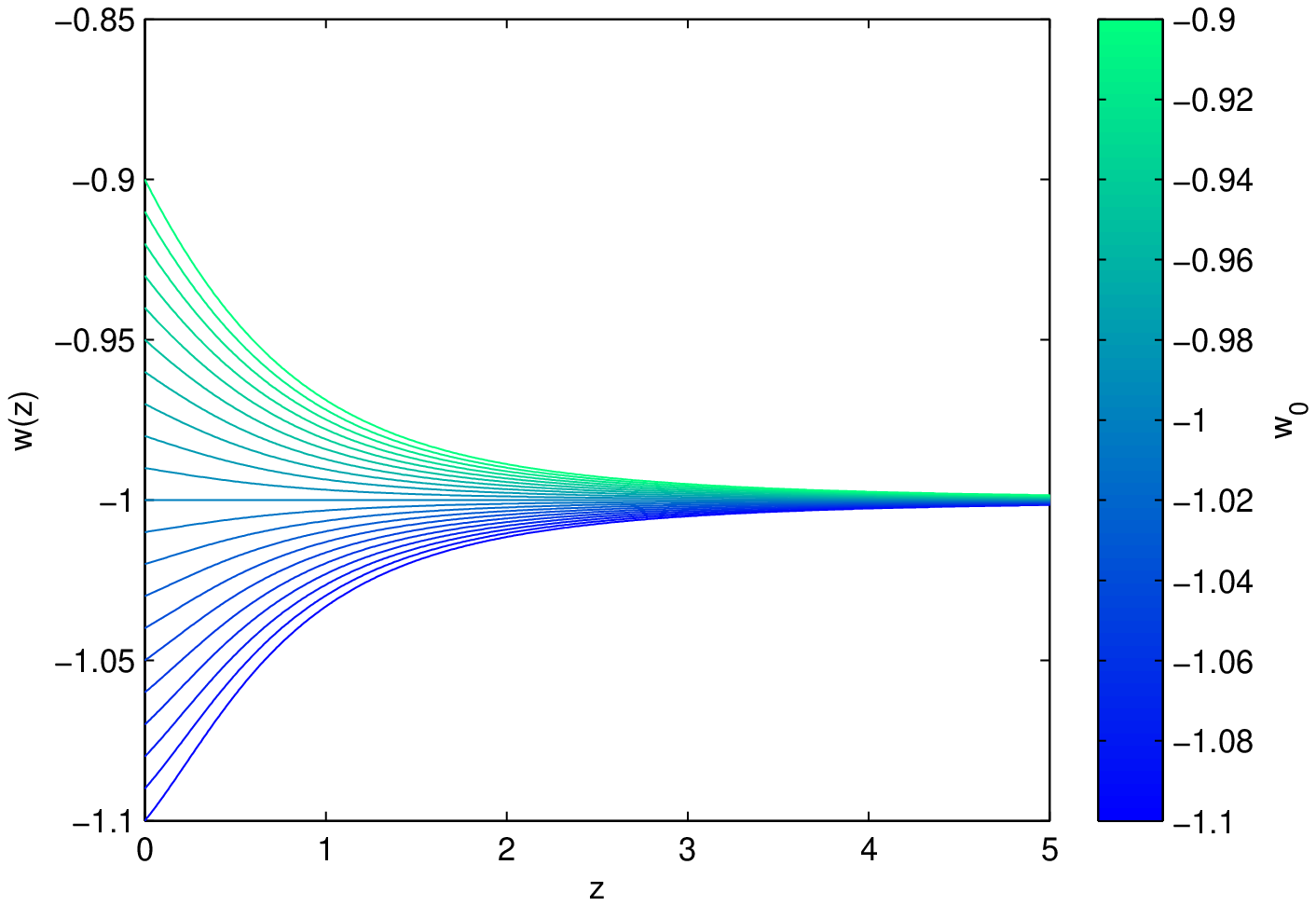}\\
\includegraphics[width=3in]{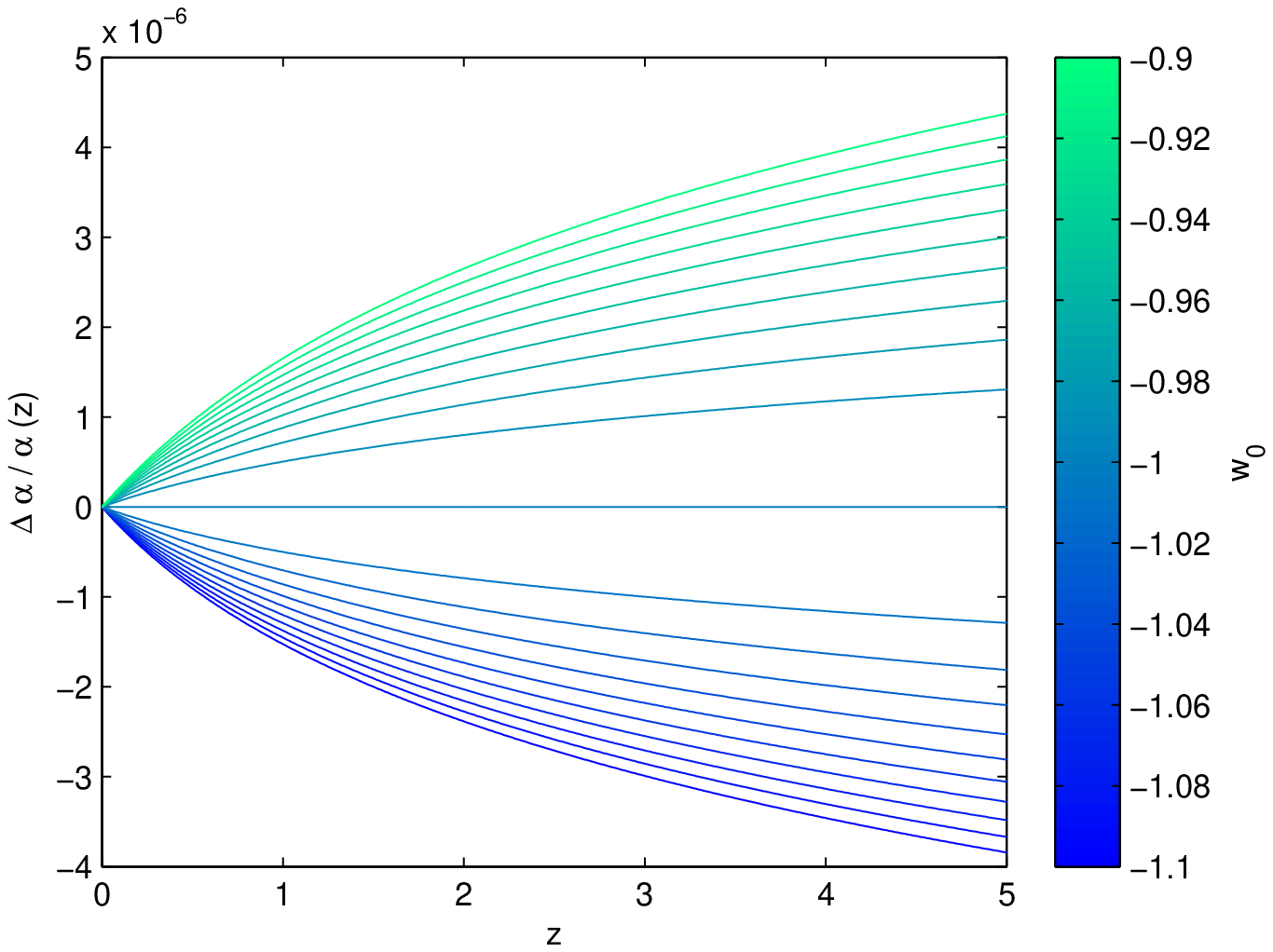}
\includegraphics[width=3in]{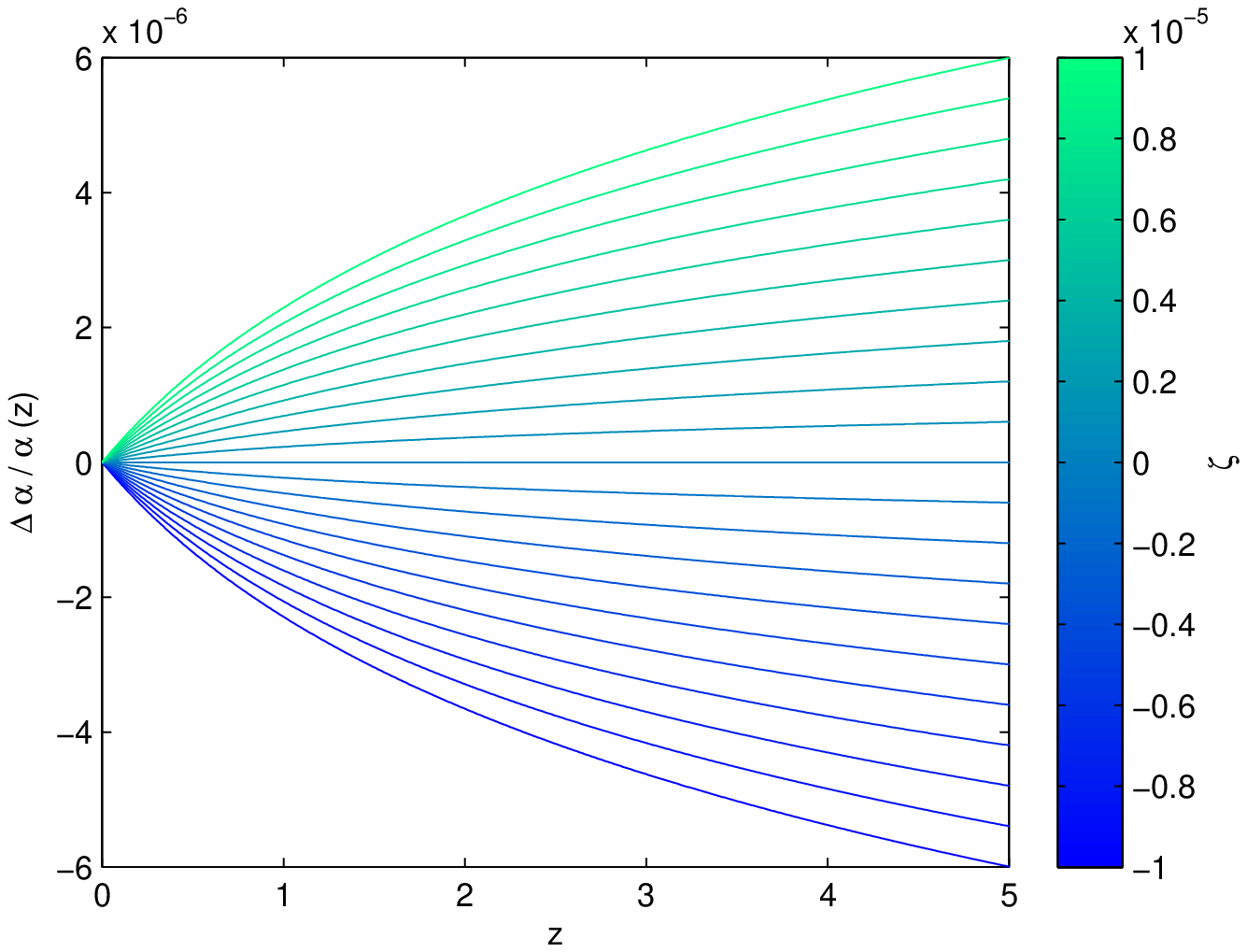}
\caption{\label{fig02}Redshift dependence of relevant parameters in the model of Slepian {\it et al.} \protect\cite{SGZ}. {\bf Top panel:} the dark energy equation of state, $w(z)$, for various choices of $w_0$. {\bf Bottom left panel:} Relative variation of the fine-structure constant, $\Delta\alpha/\alpha$, for various choices of $w_0$, assuming a coupling $\zeta=5\times10^{-6}$. {\bf Bottom right panel:} Relative variation of the fine-structure constant, for various choices of $\zeta$, assuming $w_0=-0.95$. A flat universe with $\Omega_{m0}=0.3$ has been assumed throughout.}
\end{figure}

The dark energy equation of state has the following behavior
\begin{equation}\label{darkeos}
w(z)=-1+(1+w_0)\frac{H_0^2}{H^2(z)}\,.
\end{equation}
Note that for high redshifts this always approaches -1, and it diverges from this value as the universe evolves, reaching $w_0$ today. This is therefore a parametrization for thawing models. Apart from its simplicity, this choice of parametrization is also motivated by the recent result that if physical priors are used, allowed quintessence models are mostly thawing \cite{Marsh}. Figure \ref{fig02} illustrates the behavior of the dark energy equation of state and the fine-structure constant for relevant parameter choices.

%%%%%%%%%%%%%%%%%%%%%%%%%%%%%%%%%%%%%%%%%%%%%%%%%%%%%%%%%%%%%%%%%%%%%%%%%%%%%%%%%%
%%%%%%%%%%%%%%%%%%%%%%%%%%%%%%%%%%%%%%%%%%%%%%%%%%%%%%%%%%%%%%%%%%%%%%%%%%%%%%%%%%
\begin{figure}[tbp]
\centering
\includegraphics[width=3in]{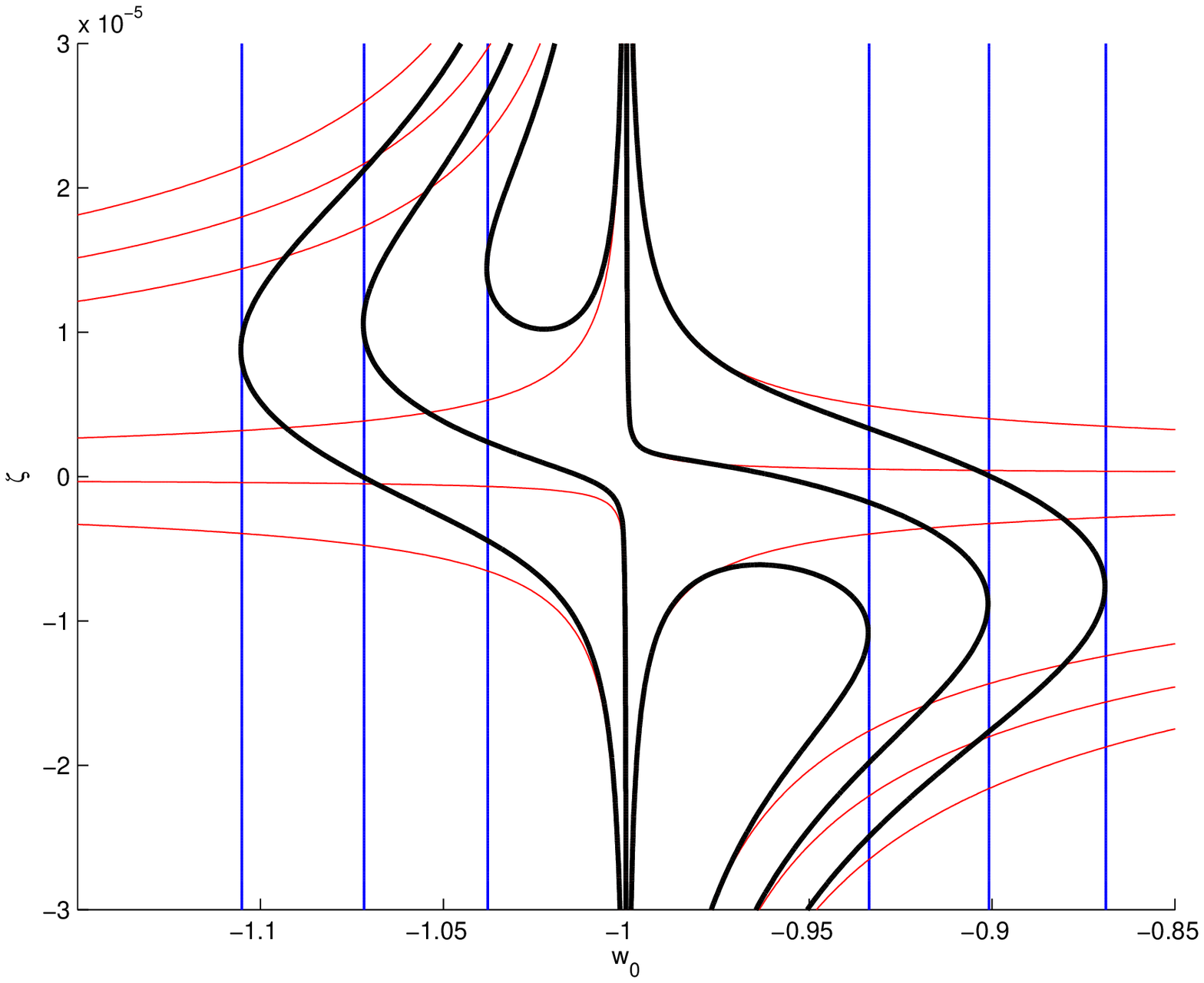}
\includegraphics[width=3in]{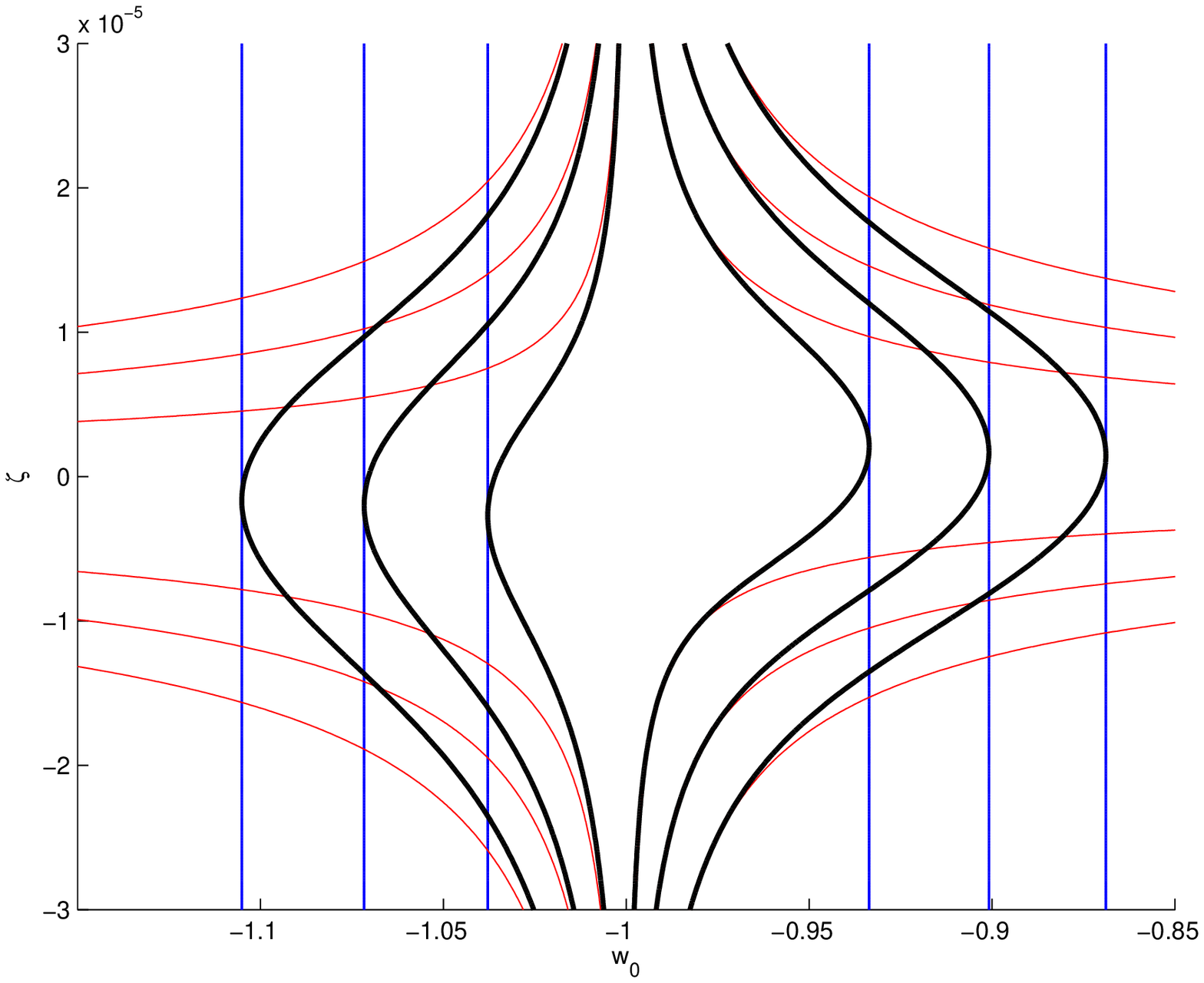}\\
\includegraphics[width=3in]{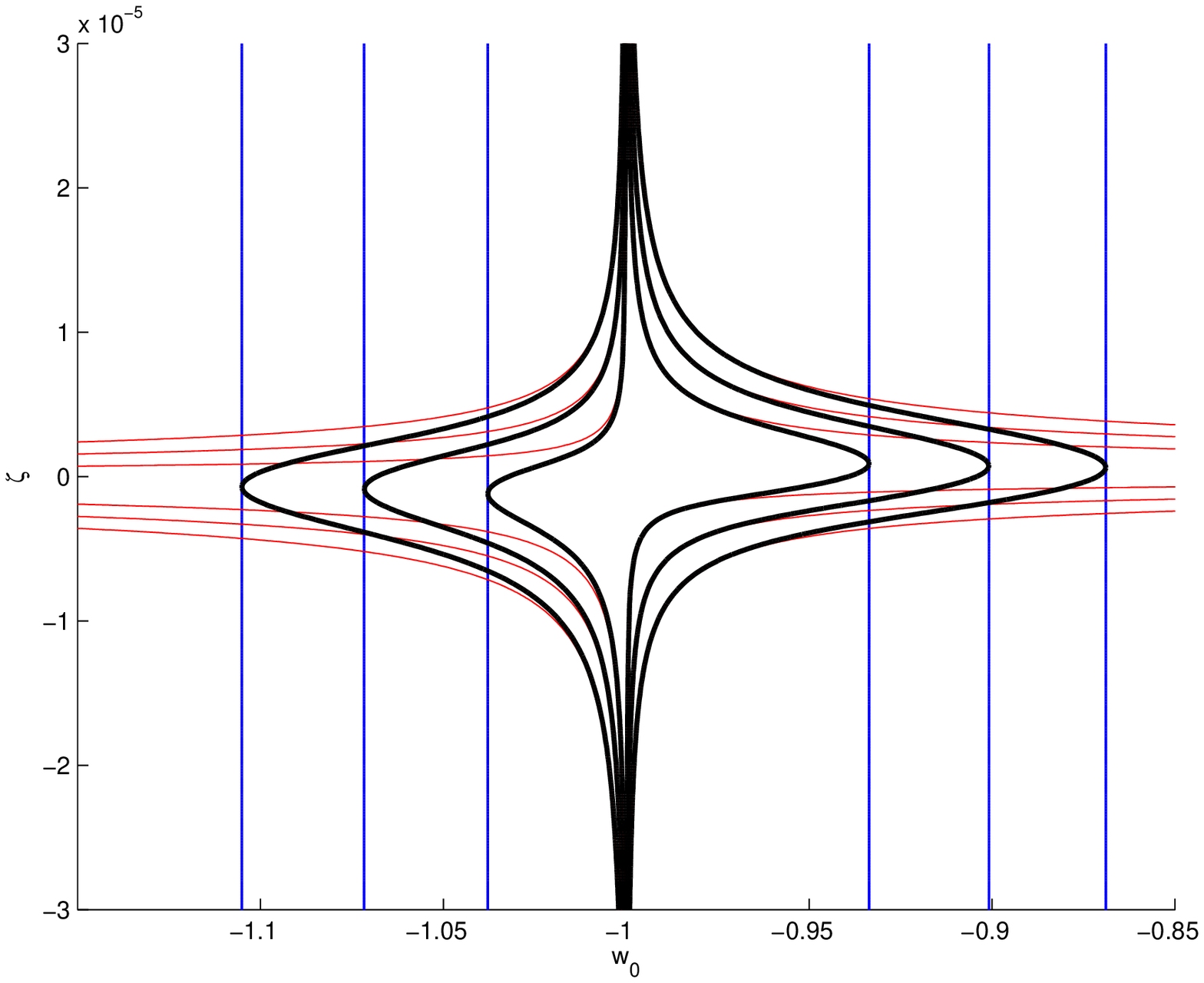}
\caption{\label{fig03}One, two and three sigma constraints on the $\zeta-w_0$ plane for the model of Slepian {\it et al.} \protect\cite{SGZ}, from Webb {\it et al.} data (top left panel), Table \protect\ref{table1} data (top right panel) and the atomic clock bound (bottom panel). In each panel the thin red lines correspond to the constraints from the astrophysical or clock data alone, the blue vertical ones correspond to the cosmological data (which constrain $w_0$ but are insensitive to $\zeta$) and the black thick lines correspond to the combined datasets.}
\end{figure}
%%%%%%%%%%%%%%%%%%%%%%%%%%%%%%%%%%%%%%%%%%%%%%%%%%%%%%%%%%%%%%%%%%%%%%%%%%%%%%%%%%

We can now proceed to comparing this class of models with the available data. Figure \ref{fig03} shows the results of this comparison, separately for the Webb {\it et al.} data (top left panel), and for the Table \protect\ref{table1} data (top right panel)---in both cases, the constraints from the astrophysical data are shown by the thin red lines. It's well known that the Webb {\it et al.} data is not consistent with the null result, and a weighted mean of the dataset yields a negative value, corresponding to a slightly smaller value of $\alpha$ in the past \cite{Webb}. In our analysis we correspondingly find a one sigma preference for a non-zero coupling $\zeta$ (with a negative sign for a canonical field, and a positive sign for a phantom field). However, the data is compatible with the null result at two sigma.

On the other hand, the Table \protect\ref{table1} data is fully compatible with the null result. It's worth noting that in the former case the reduced chi-square of the best-fit model (for the $\alpha$ data alone) is $\chi^2_{min,Webb}=1.04$, while in the latter case it is $\chi^2_{min,Table}=1.29$; this may be an indication that some of the uncertainties in the Table \protect\ref{table1} measurements have been underestimated. For comparison we also show in the bottom panel of Fig. \ref{fig03}, in the same scale as before (and also in thin red lines), the local atomic clock constraint of  Rosenband {\it et al.} \cite{Rosenband}; it is clear from the plot that this is currently more constraining than the astrophysical measurements.

\begin{figure}[tbp]
\centering
\includegraphics[width=3in]{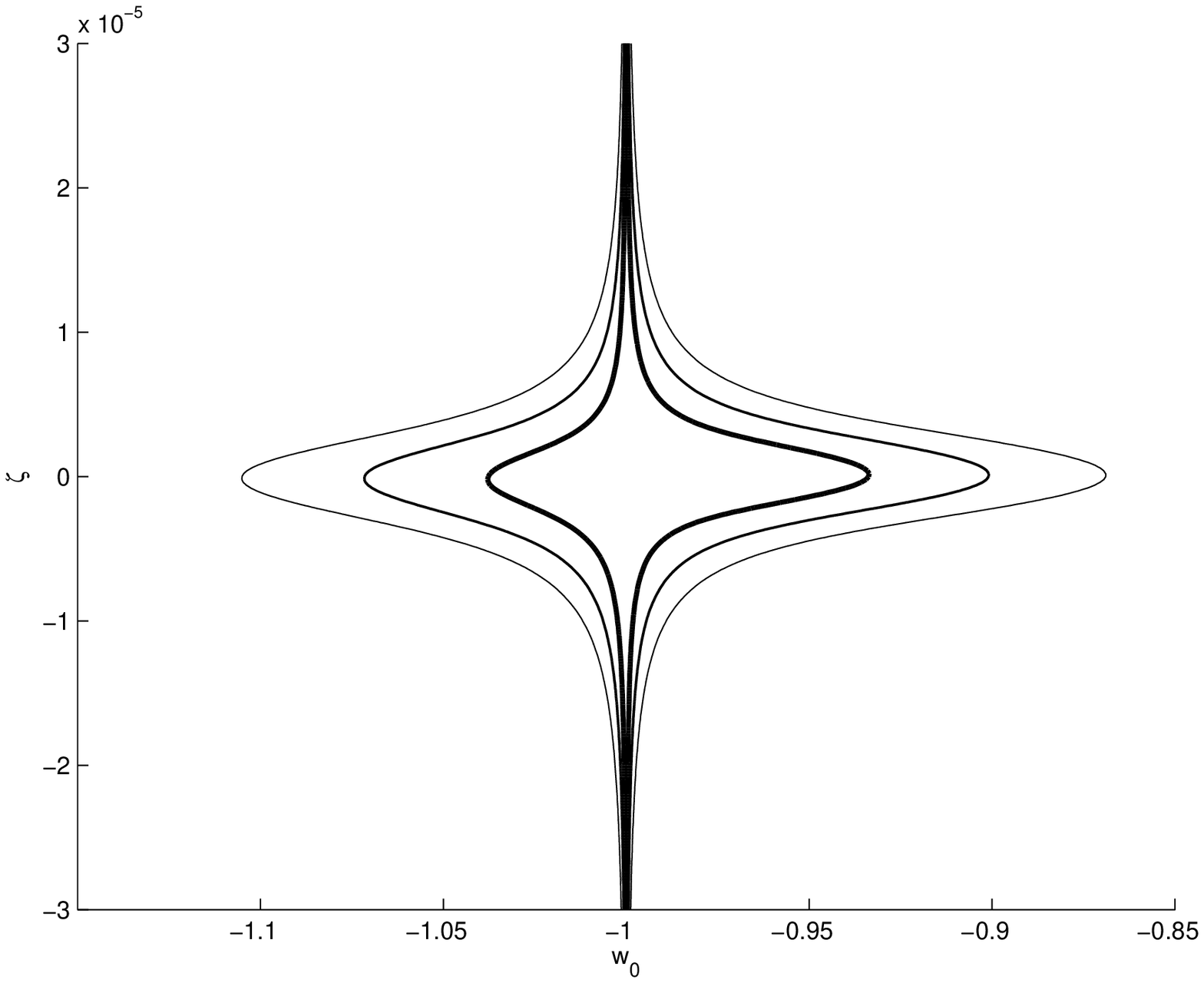}
\caption{\label{fig04}One, two and three sigma constraints on the $\zeta-w_0$ plane for the model of Slepian {\it et al.} \protect\cite{SGZ}, from the full dataset considered in our analysis: Webb {\it et al.} data plus Table \protect\ref{table1} data plus atomic clock bound plus cosmological (Type Ia supernova and Hubble parameter) data. The reduced chi-square of the best fit is $\chi^2_{min,full}=0.97$.}
\end{figure}

In Fig. \ref{fig03} the cosmological data constraints are shown by the blue vertical lines, while the combined (cosmological plus astrophysical, or cosmological plus atomic clock) constraints are shown by the thick black lines. The role of the cosmological datasets in constraining $w_0$, and thus breaking the $\zeta$-$w_0$ degeneracy, is manifest. Naturally, we can obtain tighter constraints by combining all the datasets; this is straightforward to do since the Webb {\it et al.}, Table \ref{table1} and atomic clock measurements of $\alpha$ are all independent. The results of this analysis are shown in Fig. \ref{fig04}. We note that the results of this analysis are fairly similar to those of \cite{Pinho}, where a model with a constant equation of state, $w(z)=w_0$, was assumed. The main reason for this is that the atomic clock bound, which currently dominates the constraints (as can be seen from Fig. \ref{fig03}), is only sensitive to the present value of the equation of state (ie, $w_0$ itself) and not to its evolution.

%%%%%%%%%%%%%%%%%%%%%%%%%%%%%%%%%%%%%%%%%%%%%%%%%%%%%%%%%%%%%%%%%%%%%%%%%%%%%%%%%%
%%%%%%%%%%%%%%%%%%%%%%%%%%%%%%%%%%%%%%%%%%%%%%%%%%%%%%%%%%%%%%%%%%%%%%%%%%%%%%%%%%
\begin{figure}[tbp]
\centering
\includegraphics[width=3in]{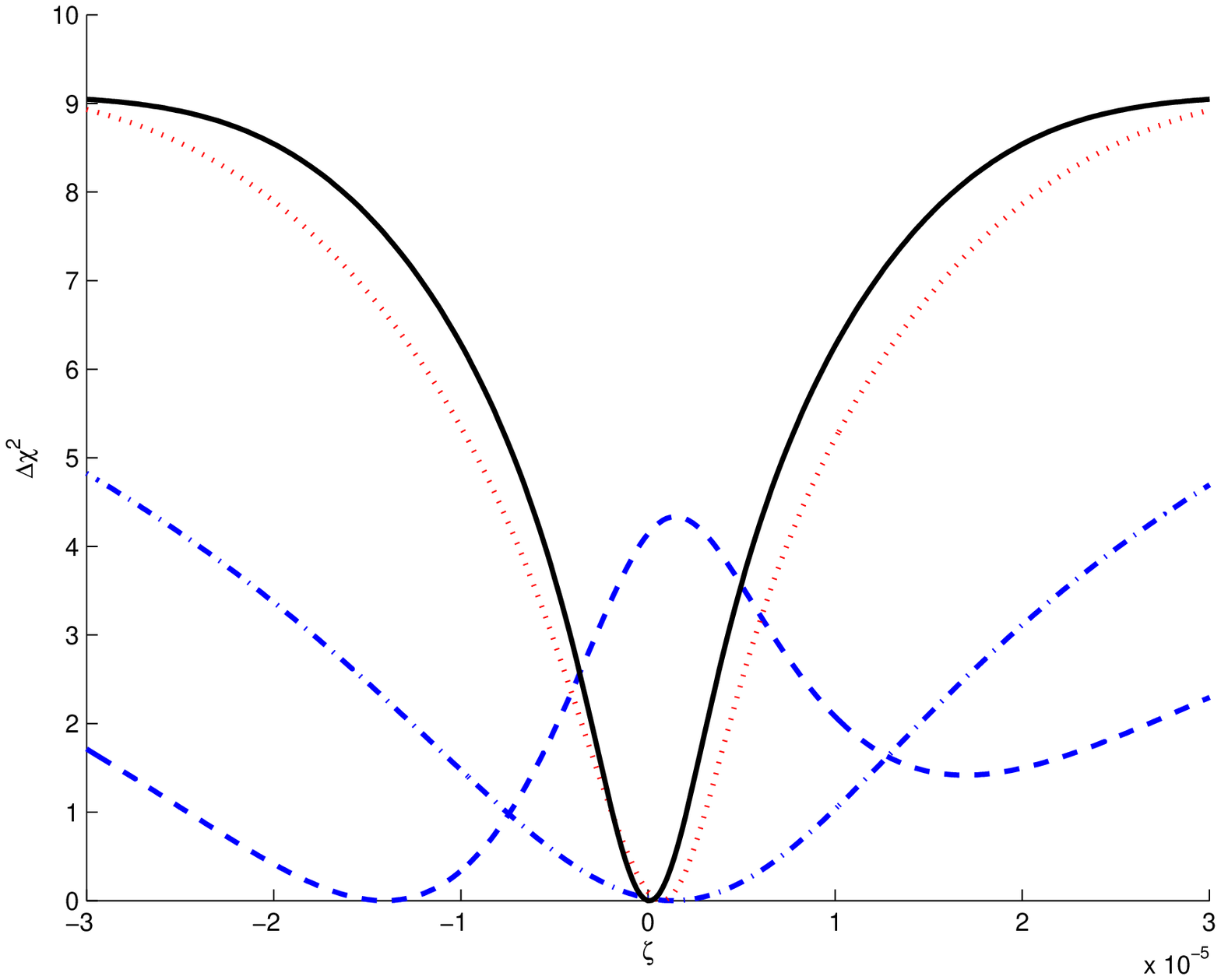}
\includegraphics[width=3in]{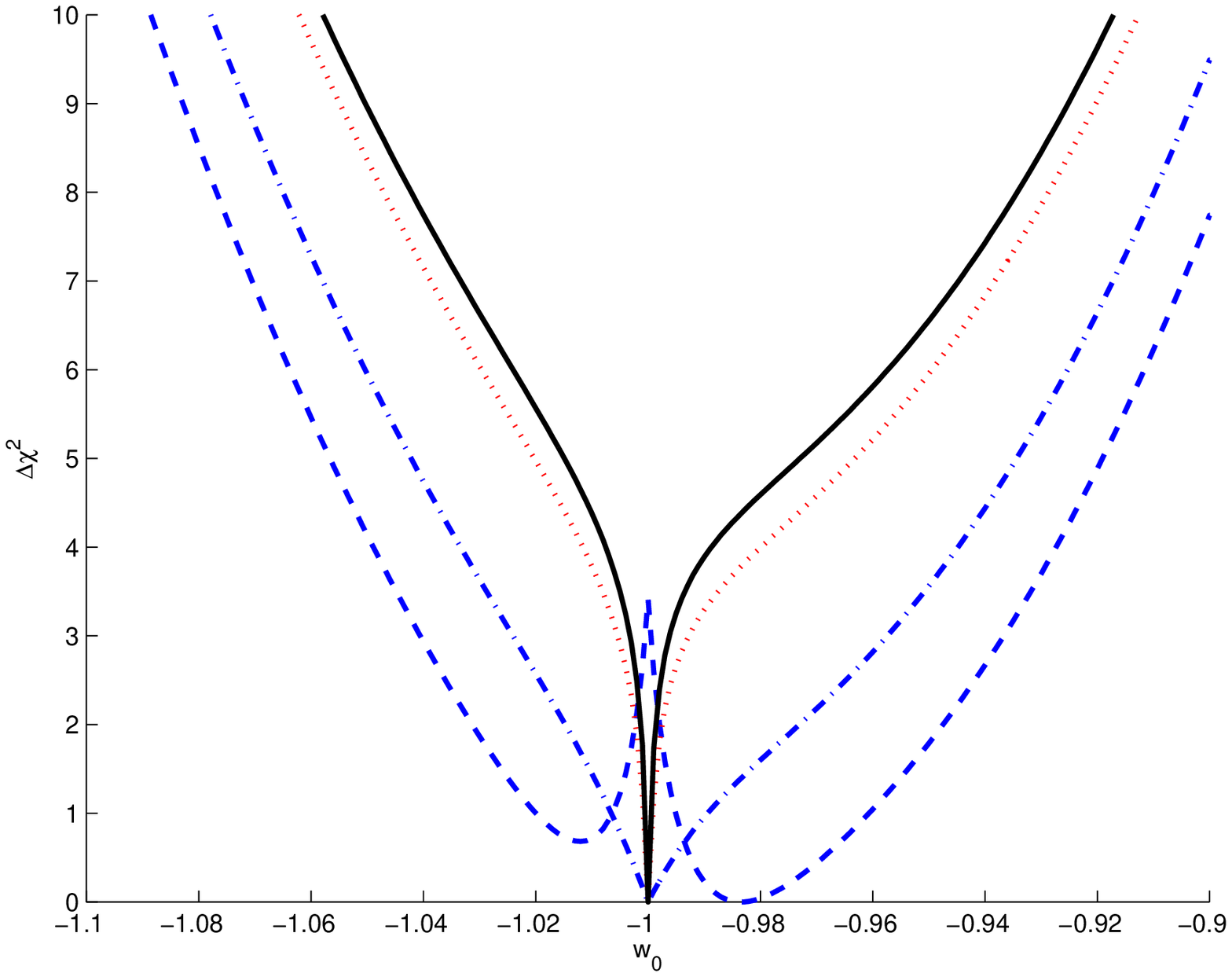}
\caption{\label{fig05}1D likelihood for $\zeta$ marginalizing over $w_0$ (left panel) and for $w_0$  marginalizing over $\zeta$ (right panel), for the model of Slepian {\it et al.} \protect\cite{SGZ}. Plotted is the value of $\Delta\chi^2=\chi^2-\chi^2_{min}$, for cosmological $+$ Webb data (blue dashed), cosmological $+$ Table \protect\ref{table1} data (blue dash-dotted), cosmological $+$ atomic clock data (red dotted) and the combination of all datasets (black solid).}
\end{figure}
%%%%%%%%%%%%%%%%%%%%%%%%%%%%%%%%%%%%%%%%%%%%%%%%%%%%%%%%%%%%%%%%%%%%%%%%%%%%%%%%%%

Finally, we can also obtain the 1D constraint on the coupling $\zeta$ by marginalizing over the dark energy equation of state $w_0$. The results of this analysis are shown in the left panel of Fig. \ref{fig05}. Again we confirm that in the case of the Webb {\it et al.} dataset there is a one-sigma preference for a non-zero coupling, while in the other cases the null result provides the best fit. The full dataset allows us to obtain a non-trivial constraint on $\zeta$. At the two-sigma ($95.4\%$) confidence level we find
\begin{equation} \label{zetaboundSGZ}
|\zeta_{\rm SGZ}|<5.6\times10^{-6}\,,
\end{equation}
which leads to a constraint on WEP violations
\begin{equation} \label{etaboundSGZ}
\eta_{\rm SGZ}<3.1\times10^{-14}\,.
\end{equation}
These constraints are very slightly weaker than those obtained in \cite{Pinho} for the constant equation of state model (cf. Eqs. \ref{zetaboundfix} and \ref{etaboundfix} respectively). Physically, the reason for this is that in a thawing model with a given $w_0$ the amount of $\alpha$ variation at a given non-zero redshift will be slightly smaller than that in a constant equation of state model with the same $w_0$. In any case, our indirect WEP bound is still stronger than the available direct bounds, cf. Eqs. \ref{boundetaT}-\ref{boundetaL}.

We can similarly obtain the 1D likelihood for $w_0$ by marginalizing over $\zeta$; this is shown on the right panel of Fig. \ref{fig05}. In this case we find at the three-sigma ($99.7\%$) confidence level
\begin{equation} \label{w0boundSGZ}
-1.05<w_0<-0.92\,,
\end{equation}
which is again slightly weaker than the one for the constant equation of state model (cf. Eq. \ref{w0boundfix}). Note that, as can be seen in the blue contours in Fig. \ref{fig03}, from our cosmology data alone we would get at three-sigma $-1.11<w_0<-0.87$, so the improvement provided by the $\alpha$ data is significant. Therefore this is nominally a very strong bound, though we note that it should be interpreted cautiously, both due to our assumptions on other cosmological parameters and also because the likelihood is clearly not Gaussian near the minimum. This raises the issue of the choice of priors---specifically, of the choice of a flat prior for $1+w_0$. We will discuss this point in the following section.

%%%%%%%%%%%%%%%%%%%%%%%%%%%%%%%%%%%%%%%%%%%%%%%%%%%%%%%%%%%%%%%%%%%%%%%%%%%%%%%%%%
\section{A class of freezing models}
\label{sec:dil}

In the previous section we provided constraints on thawing models. Here we will consider the opposite scenario: that of freezing models where the dark energy equation of state evolves towards $-1$. Our motivation here stems from the fact that in many dilaton-type models the scalar field depends logarithmically on the scale factor
\begin{equation}
{\phi}(z)\propto \log{(1+z)}\,.
\end{equation}
Therefore, for a linear gauge kinetic function as we are assuming here, it follows that in that case $\frac{\Delta\alpha}{\alpha}\propto\ln(1+z)$. We will presently calculate the condition on the dark energy equation of state for Class I models to have a such a behavior for $\alpha(z)$, but it is worth emphasizing at this point that some Class II models are also known to display such a behavior. Indeed, for runaway dilaton scenarios this behavior is even approximately true after the onset of dark energy domination (see for example \cite{DPV,NewDPV}). By contrast, in BSBM models (another Class II model) the field departs from this behavior and freezes quite abruptly at this epoch \cite{BSBM}.

One may therefore ask what kind of dark energy equation of state would lead to this behavior. From Eq. \ref{eq:dalfa} we infer that the function inside the square root therein must be a constant, that is
\begin{equation}
\Omega_{\phi}(z)[1+w(z)]=const.\,;
\end{equation}
this can be recast into the following equation
\begin{equation}
\frac{dw}{dz}=-3(1+w_0)\frac{w}{1+z}\left[\frac{1+w}{1+w_0}-\Omega_{\phi0}\right]\label{eq:diff}
\end{equation}
Note that the initial condition for the first derivative is
\begin{equation}
\left[\frac{dw}{dz}\right]_{z=0}=-3\Omega_m w_0(1+w_0)\,,
\end{equation}
and for the second one we could also write
\begin{equation}
\left[\frac{d^2w}{dz^2}\right]_{z=0}=3\Omega_m w_0(1+w_0)[1+3w_0+3\Omega_m(1+w_0)]\,,
\end{equation}
so $w'\sim3\Omega_m(1+w_0)$ and $w''\sim6\Omega_m(1+w_0)$ near the $\Lambda$CDM limit.

%%%%%%%%%%%%%%%%%%%%%%%%%%%%%%%%%%%%%%%%%%%%%%%%%%%%%%%%%%%%%%%%%%%%%%%%%%%%%%%%%%
\begin{figure}[tbp]
\centering
\includegraphics[width=3in]{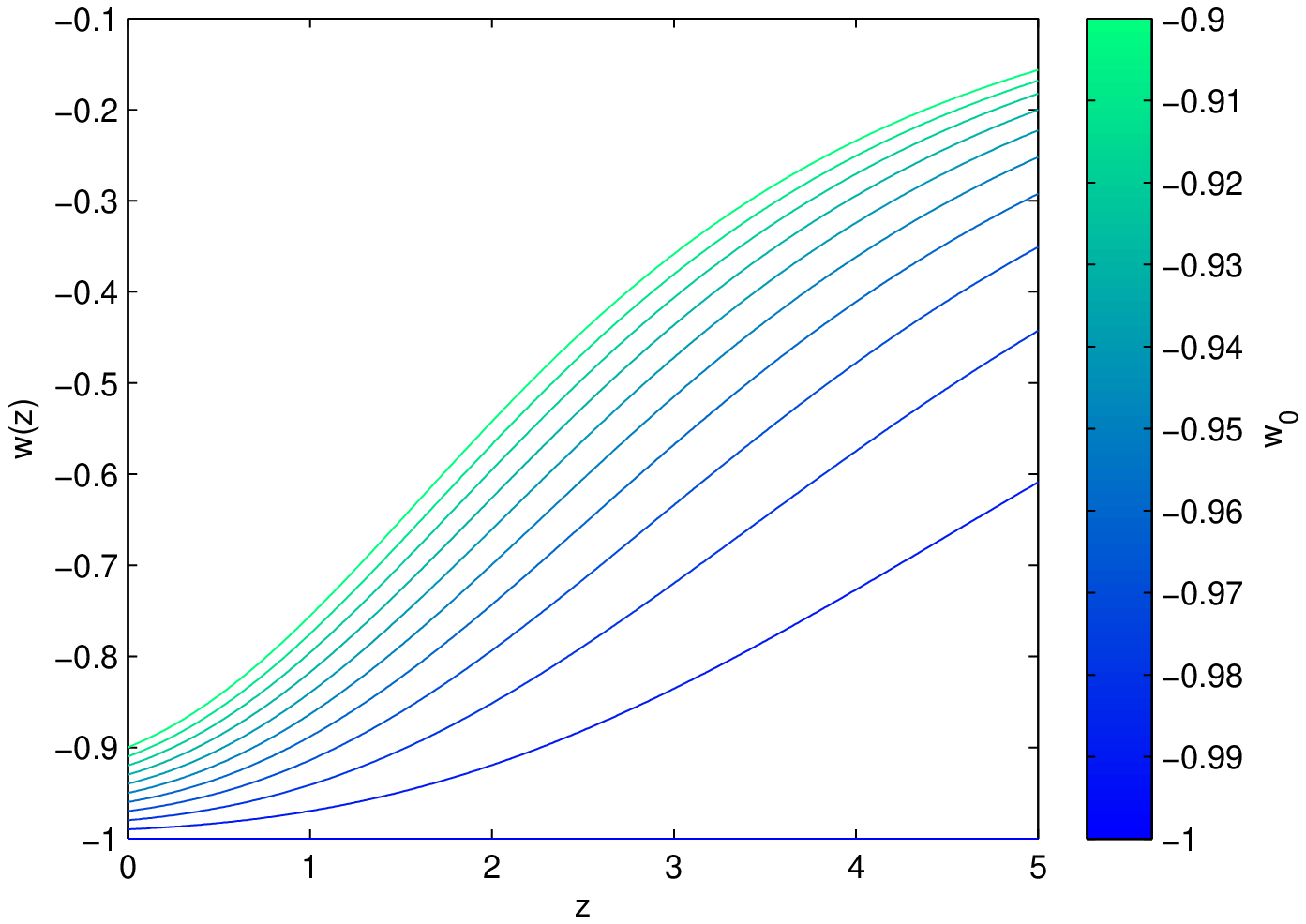}\\
\includegraphics[width=3in]{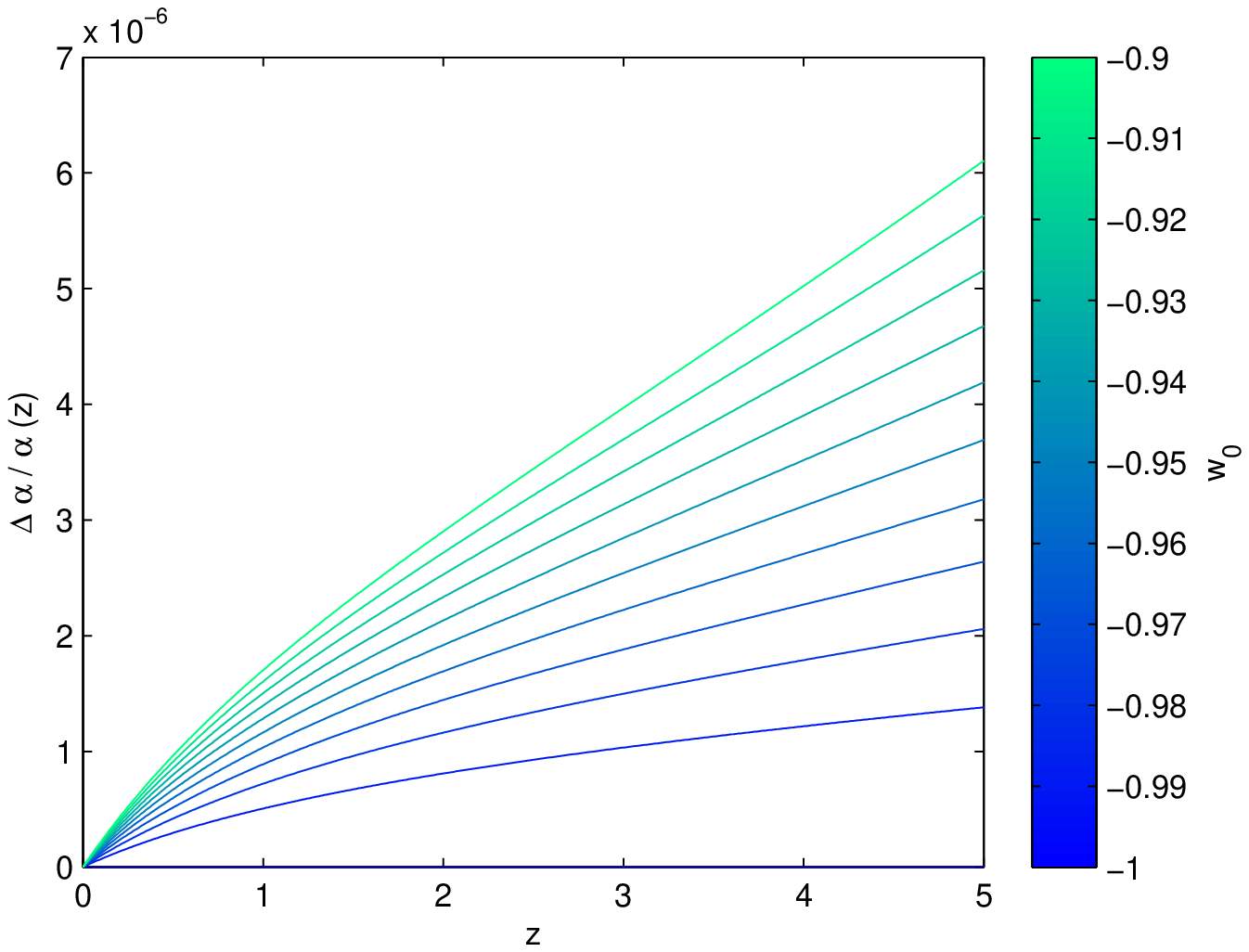}
\includegraphics[width=3in]{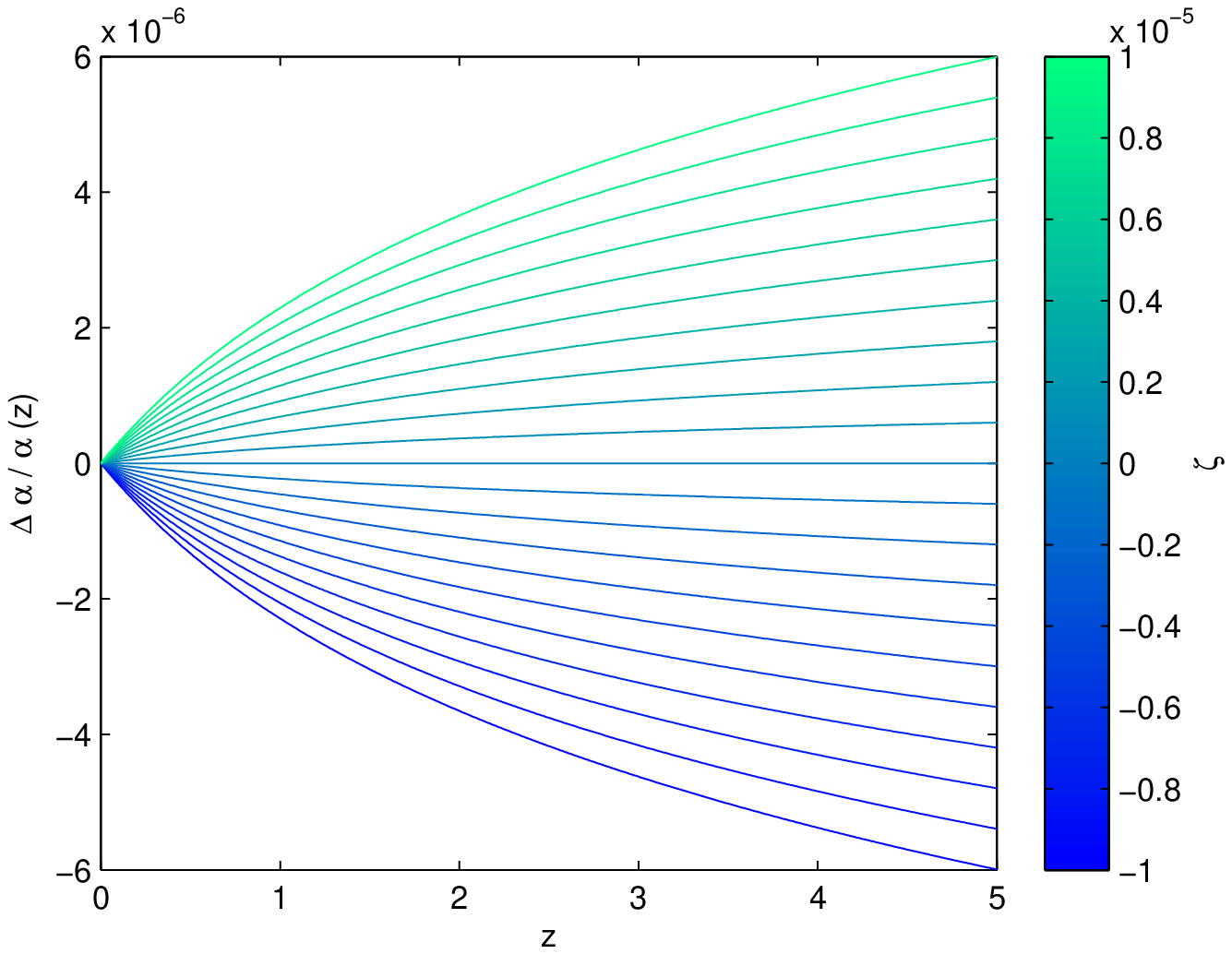}
\caption{\label{fig06}Redshift dependence of relevant parameters in the dilaton-like class of models. {\bf Top panel:} the dark energy equation of state, $w(z)$, for various choices of $w_0$. {\bf Bottom left panel:} Relative variation of the fine-structure constant, $\Delta\alpha/\alpha$, for various choices of $w_0$, assuming a coupling $\zeta=5\times10^{-6}$. {\bf Bottom right panel:} Relative variation of the fine-structure constant, for various choices of $\zeta$, assuming $w_0=-0.95$. A flat universe with $\Omega_{m0}=0.3$ has been assumed throughout.}
\end{figure}
%%%%%%%%%%%%%%%%%%%%%%%%%%%%%%%%%%%%%%%%%%%%%%%%%%%%%%%%%%%%%%%%%%%%%%%%%%%%%%%%%%

The above equation can be easily integrated, leading to the solution
\begin{equation}\label{darkeos2}
w(z)=\frac{[1-\Omega_\phi(1+w_0)]w_0}{\Omega_m(1+w_0)(1+z)^{3[1-\Omega_\phi(1+w_0)]}-w_0}\,,
\end{equation}
where as usual we are assuming that $\Omega_m+\Omega_\phi=1$. An analogous solution was obtained, in a different context, in \cite{Nunes}.

The Friedmann equation in this case has the explicit form
\begin{equation}\label{newsqrt2}
\frac{H^2(z)}{H^2_0}= \Omega_{m}(1+z)^{3}+\frac{\Omega_{\phi}}{\Omega_m(1+w_0)-w_0}\left[\Omega_m(1+w_0)(1+z)^3-w_0(1+z)^{3\Omega_\phi(1+w_0)}\right]\,,
\end{equation}
and naturally the evolution of $\alpha$ is given by
\begin{equation}
\frac{\Delta\alpha}{\alpha}(z)=\zeta\, \sqrt{3\Omega_\phi(1+w_0)}\, \ln{(1+z)}\,.
\end{equation}
Figure \ref{fig06} shows the behavior of the dark energy equation of state and the fine-structure constant in this class of models for relevant parameter choices.

%%%%%%%%%%%%%%%%%%%%%%%%%%%%%%%%%%%%%%%%%%%%%%%%%%%%%%%%%%%%%%%%%%%%%%%%%%%%%%%%%%
\begin{figure}[tbp]
\centering
\includegraphics[width=3in]{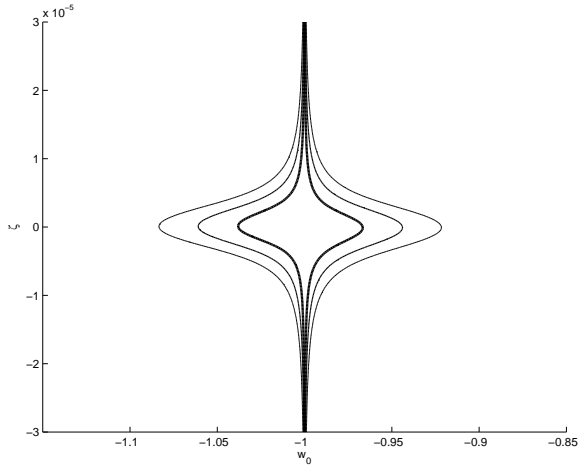}
\caption{\label{fig07}One, two and three sigma constraints on the $\zeta-w_0$ plane for dilaton-like models, from the full dataset considered in our analysis: Webb {\it et al.} data plus Table \protect\ref{table1} data plus atomic clock bound plus cosmological (Type Ia supernova and Hubble parameter) data. The reduced chi-square of the best fit is $\chi^2_{min,full}=0.97$. For this analysis we phenomenologically allowed for values of $w_0<-1$. Compare to the analogous plot for thawing models, Fig. \protect\ref{fig04}.}
\end{figure}
%%%%%%%%%%%%%%%%%%%%%%%%%%%%%%%%%%%%%%%%%%%%%%%%%%%%%%%%%%%%%%%%%%%%%%%%%%%%%%%%%%

We could treat this parametrization phenomenologically, allow for values of $w_0<-1$ (with a flat prior on $1+w_0$), and fit it to our datasets as was done in the previous section. This would lead to slightly tighter constraints, for the reason already explained: in a freezing model with a given $w_0$ the amount of $\alpha$ variation at a given non-zero redshift will be slightly larger than that in a constant equation of state model with the same $w_0$. The full dataset constraints on the $\zeta-w_0$ plane are shown in Fig. \ref{fig07}, which should be compared to the analogous plot for thawing models, Fig. \protect\ref{fig04}. As for 1D marginalized likelihoods, these would now become $|\zeta|<4.6\times10^{-6}$ at the two-sigma confidence level and $-1.04<w_0<-0.96$ at the three sigma confidence level.

However, the above analysis may be too simplistic, since in this case one physically expects that $w_0\ge-1$. We therefore discard the phantom part of this parameter, and use this model as a testbed for the effects of the choice of priors: instead of the flat prior on $1+w_0$ we have used up to this point, we will now assume a logarithmic one.

%%%%%%%%%%%%%%%%%%%%%%%%%%%%%%%%%%%%%%%%%%%%%%%%%%%%%%%%%%%%%%%%%%%%%%%%%%%%%%%%%%
\begin{figure}[tbp]
\centering
\includegraphics[width=3in]{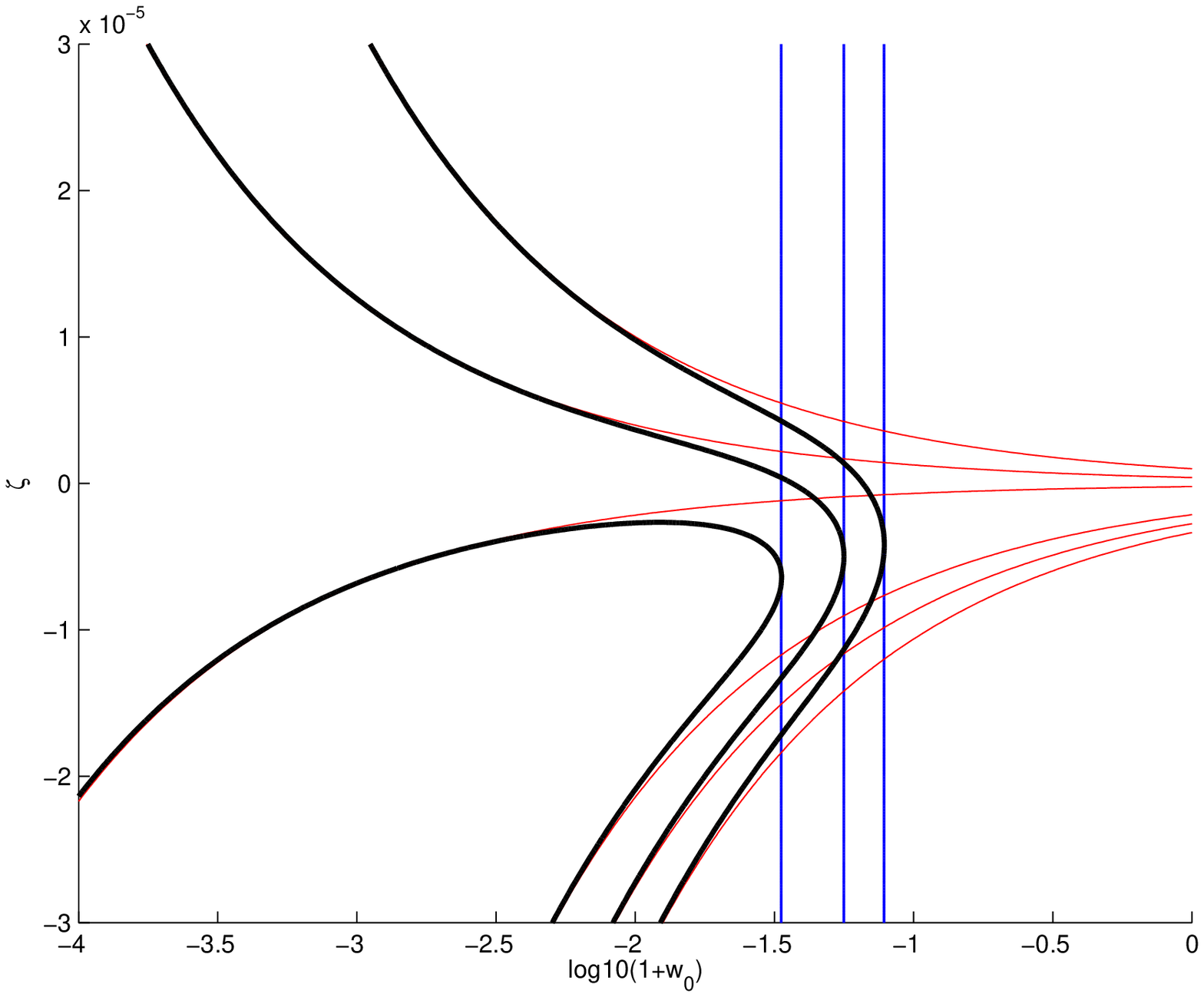}
\includegraphics[width=3in]{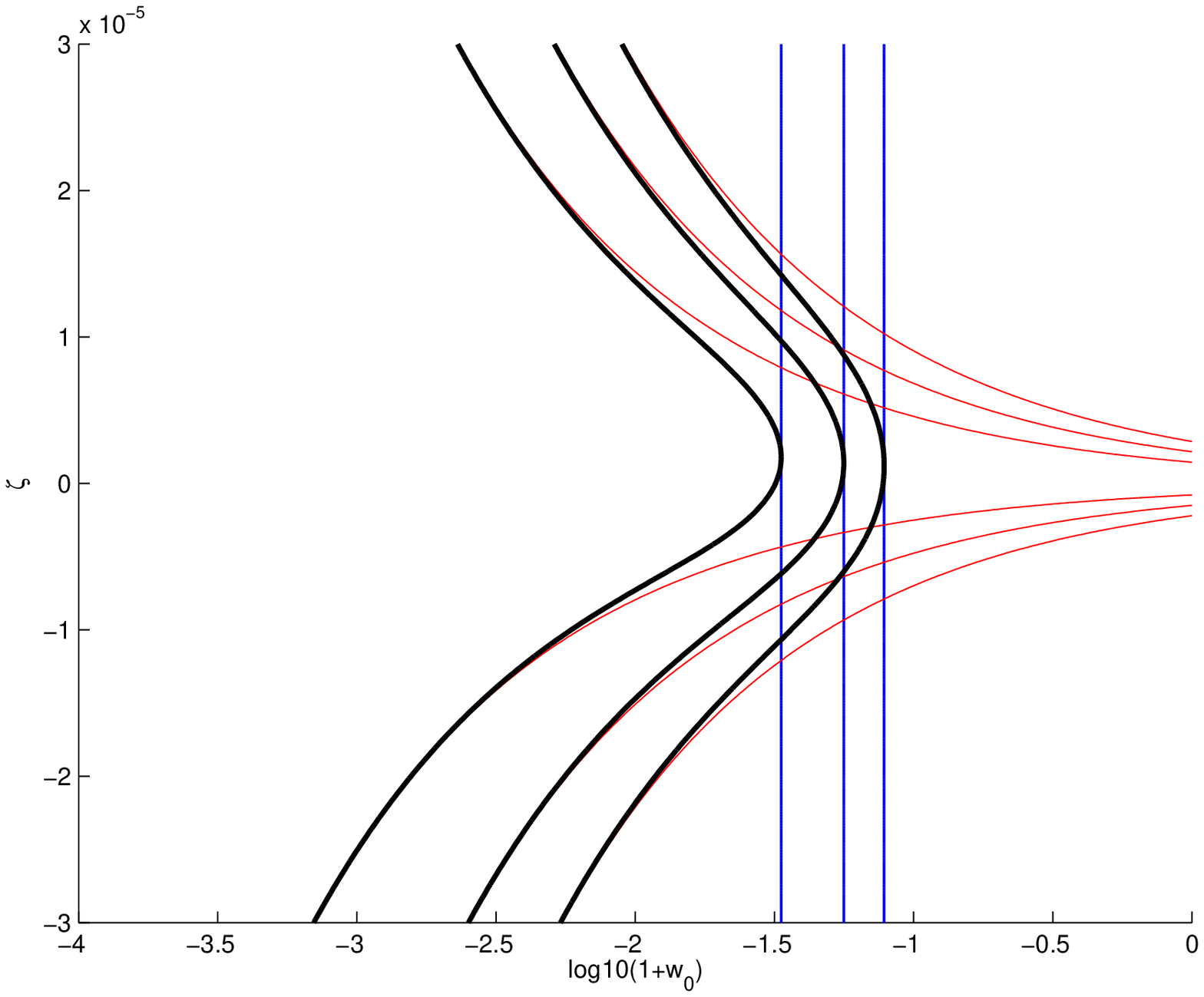}\\
\includegraphics[width=3in]{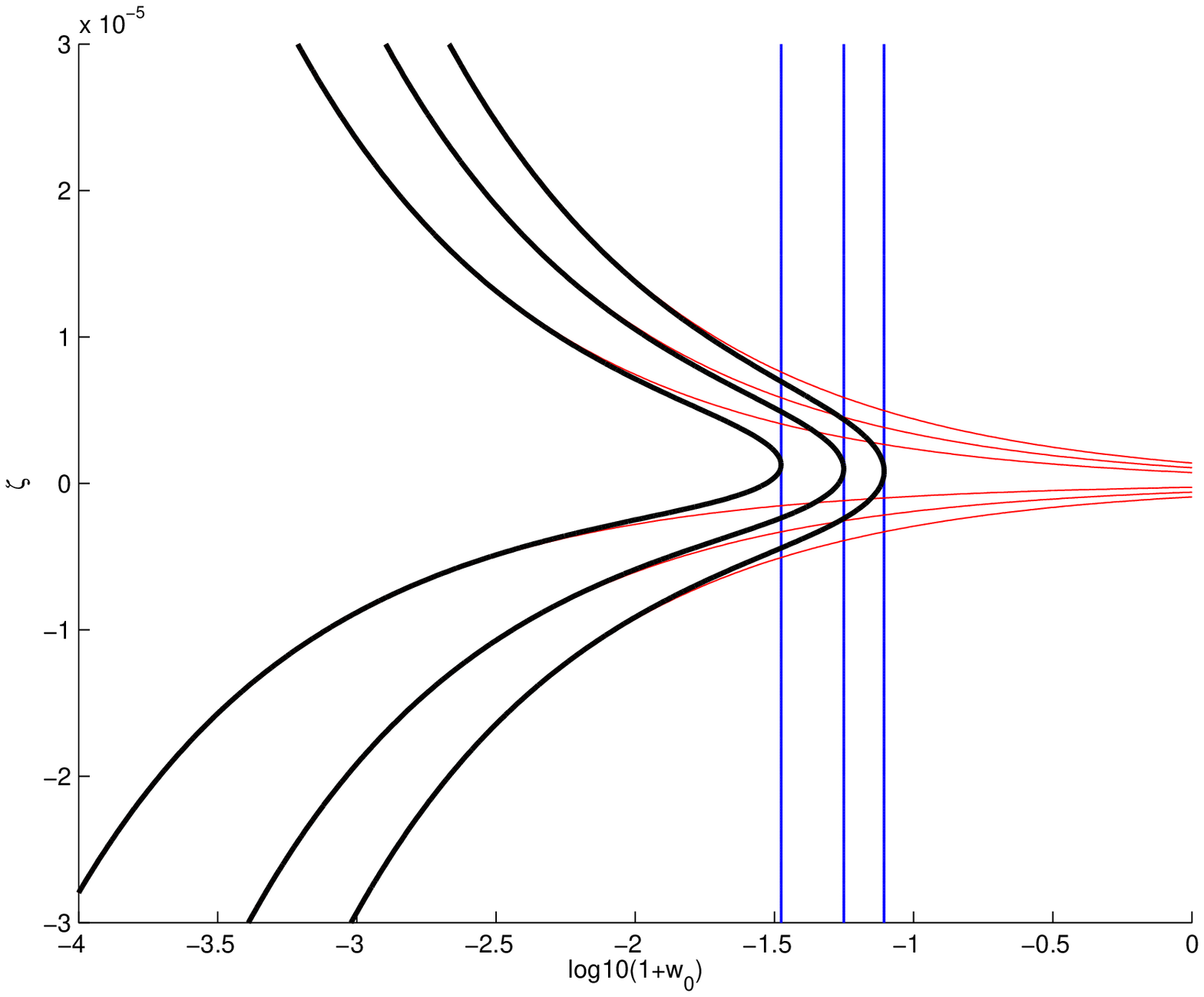}
\caption{\label{fig08}One, two and three sigma constraints on the $\zeta-w_0$ plane for dilaton-like models with a logarithmic prior on the dark energy equation of state, from Webb {\it et al.} data (top left panel), Table \protect\ref{table1} data (top right panel) and the atomic clock bound (bottom panel). In each panel the thin red lines correspond to the constraints from the astrophysical or clock data alone, the blue vertical ones correspond to the cosmological data (which constrain $w_0$ but are insensitive to $\zeta$) and the black thick lines correspond to the combined datasets.}
\end{figure}
%%%%%%%%%%%%%%%%%%%%%%%%%%%%%%%%%%%%%%%%%%%%%%%%%%%%%%%%%%%%%%%%%%%%%%%%%%%%%%%%%%

Figure \ref{fig08} shows the results of this analysis for the individual $\alpha$  datasets, alone and in combination of the cosmological data. Again we see that for the Webb {\it et al.} data there is a one-sigma preference for a non-zero value of the coupling while the data of Table \protect\ref{table1} is compatible with the null result. The reduced chi-square values for the best-fit model, for the $\alpha$ data alone, are now $\chi^2_{min,Webb}=1.05$ and $\chi^2_{min,Table}=1.28$ respectively.

\begin{figure}[tbp]
\centering
\includegraphics[width=3in]{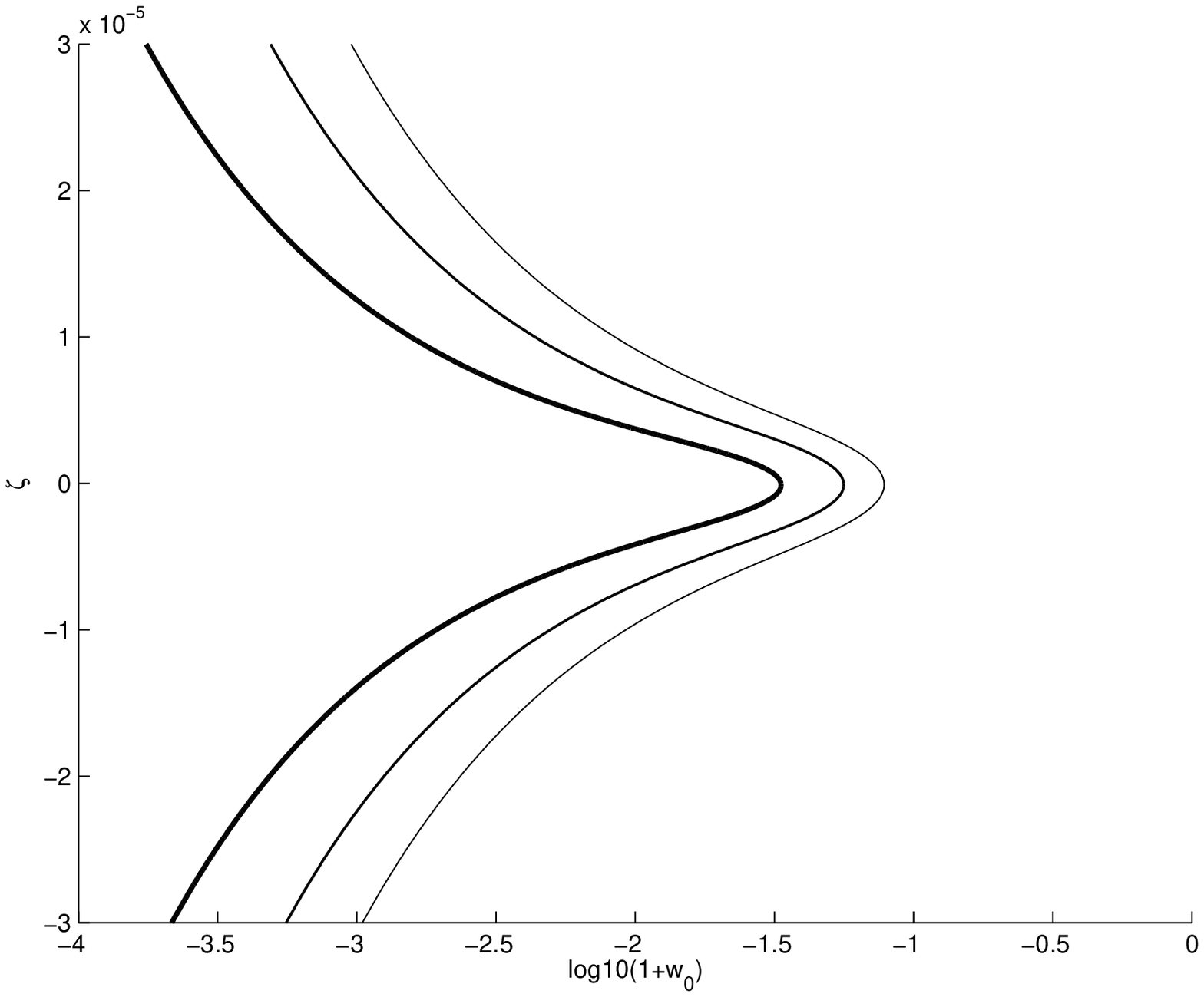}
\caption{\label{fig09}One, two and three sigma constraints on the $\zeta-w_0$ plane for dilaton-like models with a logarithmic prior on the dark energy equation of state, from the full dataset considered in our analysis: Webb {\it et al.} data plus Table \protect\ref{table1} data plus atomic clock bound plus cosmological (Type Ia supernova and Hubble parameter) data. The reduced chi-square of the best fit is $\chi^2_{min,full}=0.97$.}
\end{figure}

These plots, as well as the corresponding constraints from the combined datasets shown in Fig. \ref{fig09}, also make it clear that for a value of $w_0$ sufficiently close to $w_0=-1$ any value of the coupling would in principle be allowed---although in practice the local WEP constraints should of course be satisfied. In principle, and given the form of Eqs. \ref{eq:dalfa}-\ref{eq:dalfa2}, exactly the same would also happen in the orthogonal direction (for a sufficiently small $\zeta$ any $w_0$ would be allowed), but in practice this is prevented by the strong priors on $w_0$ coming from the cosmological datasets.

%%%%%%%%%%%%%%%%%%%%%%%%%%%%%%%%%%%%%%%%%%%%%%%%%%%%%%%%%%%%%%%%%%%%%%%%%%%%%%%%%%
%%%%%%%%%%%%%%%%%%%%%%%%%%%%%%%%%%%%%%%%%%%%%%%%%%%%%%%%%%%%%%%%%%%%%%%%%%%%%%%%%%
\begin{figure}[tbp]
\centering
\includegraphics[width=3in]{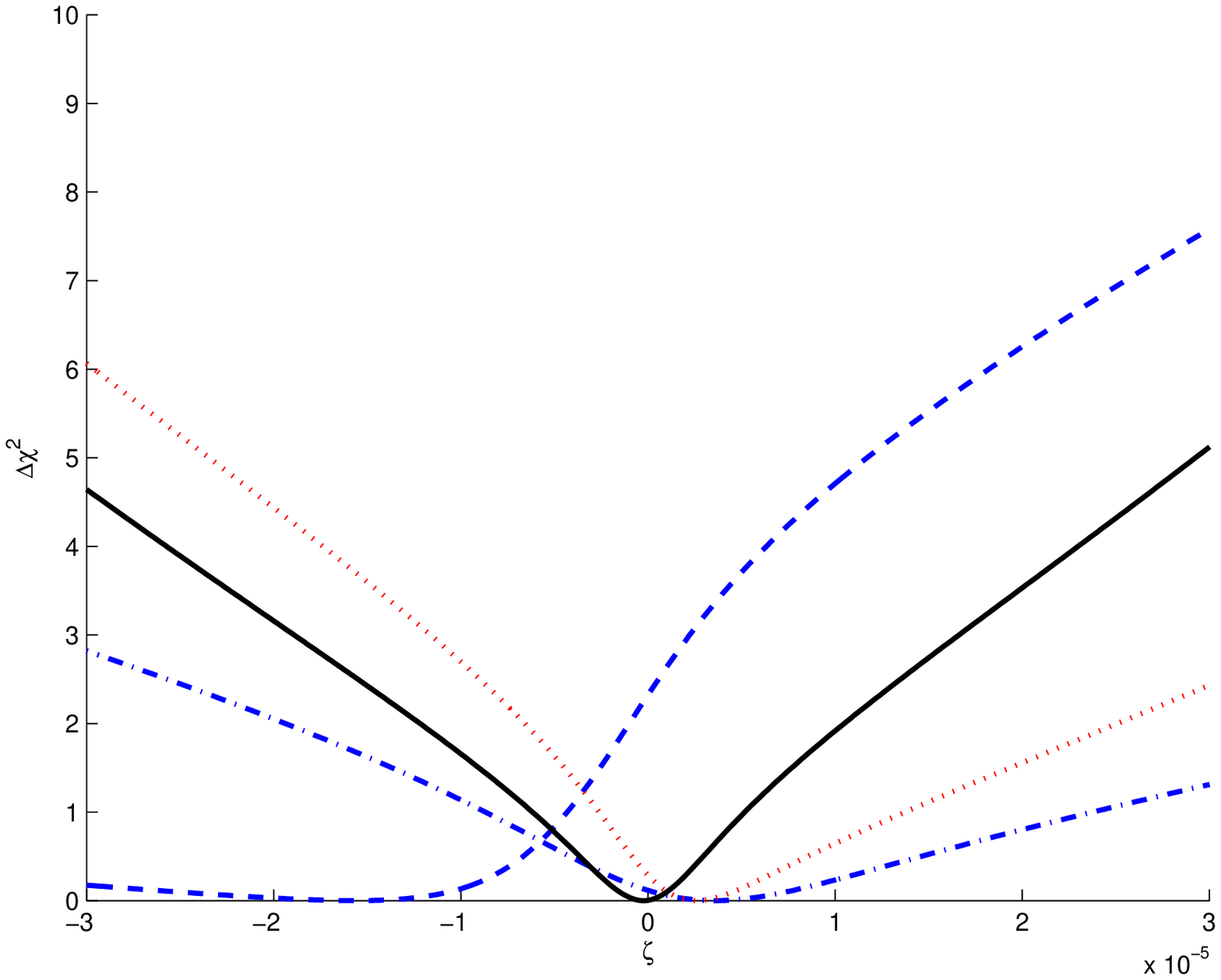}
\includegraphics[width=3in]{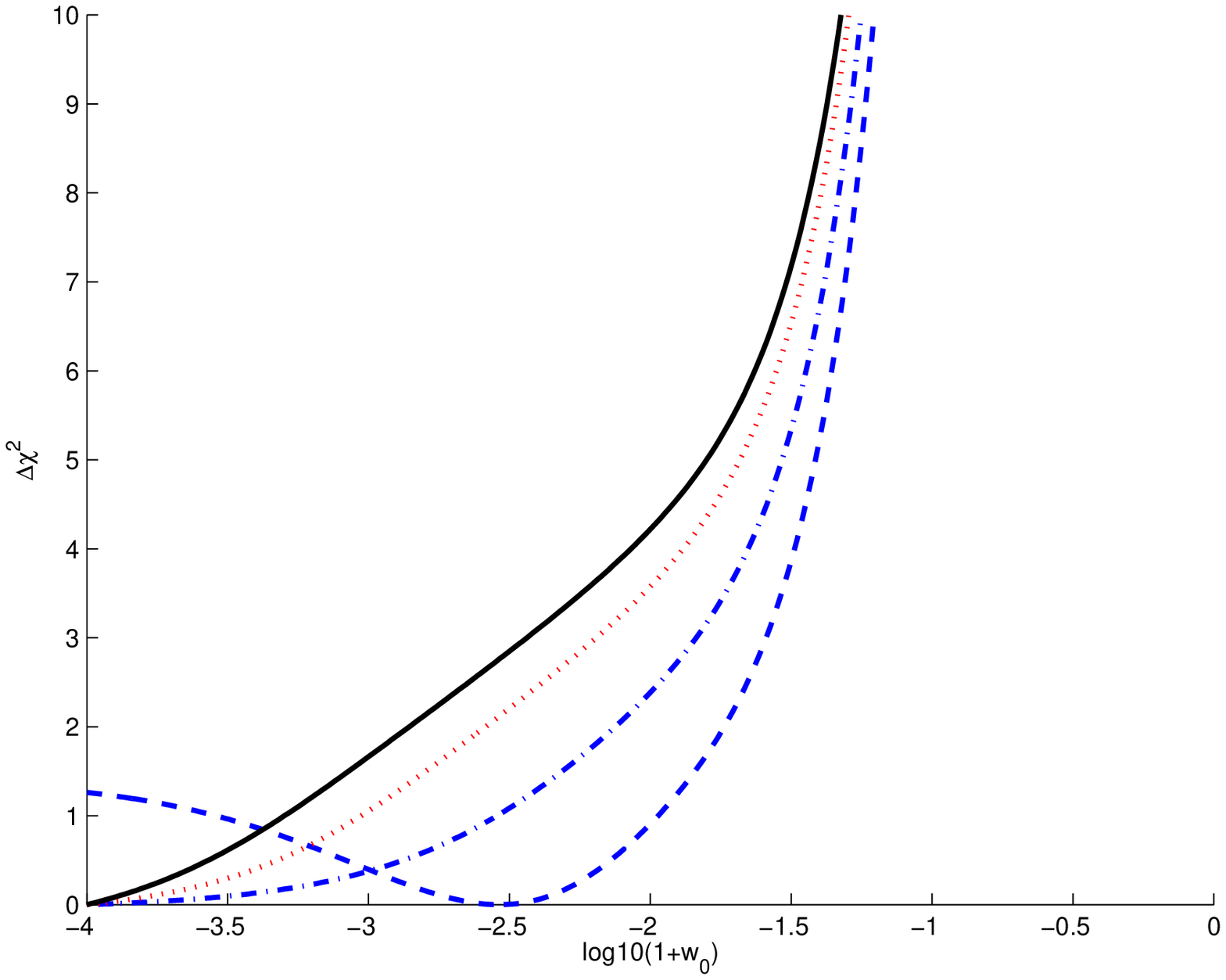}
\caption{\label{fig10}1D likelihood for $\zeta$ marginalizing over $w_0$ (left panel) and for $w_0$  marginalizing over $\zeta$ (right panel), for dilaton-like models with a logarithmic prior on the dark energy equation of state. Plotted is the value of $\Delta\chi^2=\chi^2-\chi^2_{min}$, for cosmological $+$ Webb data (blue dashed), cosmological $+$ Table \protect\ref{table1} data (blue dash-dotted), cosmological $+$ atomic clock data (red dotted) and the combination of all datasets (black solid).}
\end{figure}
%%%%%%%%%%%%%%%%%%%%%%%%%%%%%%%%%%%%%%%%%%%%%%%%%%%%%%%%%%%%%%%%%%%%%%%%%%%%%%%%%%

The 1D marginalized likelihoods for $\zeta$ and $w_0$ are depicted in Fig. \ref{fig10}. The bounds on the dark energy equation of state are unaffected, as we again find
\begin{equation} \label{w0boundDIL}
w_0<-0.96\,
\end{equation}
at the three sigma confidence level (as compared to $w_0<-0.92$ from the cosmology data alone, cf. Fig \ref{fig08}), while those on the coupling are somewhat weakened. At the one-sigma confidence level we now have
\begin{equation} \label{zetaboundDIL1}
|\zeta_{\rm DIL}|<6\times10^{-6}\,,
\end{equation}
while at the two-sigma level
\begin{equation} \label{zetaboundDIL2}
|\zeta_{\rm DIL}|<2.5\times10^{-5}\,;
\end{equation}
translating these into WEP bounds, our one sigma constraint is still stronger than the direct bounds, cf. Eqs. \ref{boundetaT}-\ref{boundetaL}: we have at the one-sigma confidence level
\begin{equation} \label{etaboundDIL2}
|\eta_{\rm DIL}|<4\times10^{-14}\,.
\end{equation}
We thus conclude that although our constraints do exhibit some model dependence (both in terms of the class of models being assumed and in terms of the underlying priors), they are generically competitive with other existing tests of these models.

As previously mentioned, in this class of models the logarithmic dependence of the scalar field, and therefore that of $\alpha$, is assumed to persist until the present day. Implicitly this assumes that the field is not significantly slowed down by the onset of dark energy domination. It is instructive to compare our constraints with those obtained in models where the field is significantly slowed down at this epoch. The BSBM class of models is the simplest toy model where this occurs. It can be shown that in this case the drift of the fine-structure constant at the present day is to a good approximation given by \cite{BSBM2}
\begin{equation} \label{BSBMdrift}
\frac{1}{H_0}\frac{\dot\alpha}{\alpha} = -\zeta_{\rm BSBM}\frac{\Omega_{m0}}{2\pi}e^{-H_0t_0}\,.
\end{equation}
Unlike Eq. \ref{clocks2} which applies to the models we have been studying, this one does not depend on the dark energy equation of state, the reason being that in these models the $\alpha$ variation is crafted onto the theory in such a way that it does not significantly affect the background dynamics (and therefore a cosmological constant is still assumed to provide the dark energy). In this case the Rosenband {\it et al.} bound given by Eq. \ref{clocks} leads to
\begin{equation} \label{zetaboundbsbm2}
\zeta_{\rm BSBM}=(0.8\pm1.2)\times10^{-4}\,
\end{equation}
at the one-sigma confidence level. As expected, for the BSBM scenario the fact that the field dynamics is strongly damped close to the present time implies that the atomic clocks bound is much weaker than the one obtained in \cite{Leal} from the astrophysical measurements of $\alpha$, cf. Eq. \ref{zetaboundbsbm}.

%%%%%%%%%%%%%%%%%%%%%%%%%%%%%%%%%%%%%%%%%%%%%%%%%%%%%%%%%%%%%%%%%%%%%%%%%%%%%%%%%%
\section{Conclusions and outlook}
\label{sec:outlook}

In this work we used a combination of astrophysical spectroscopy and local laboratory tests of the stability of the fine-structure constant $\alpha$, complemented by background cosmological datesets, to constrain the simplest examples of Class I dynamical dark energy models (in the terminology of \cite{cjmGRG}),  where the same degree of freedom is responsible for both the dark energy and a variation of $\alpha$. In these models the behavior of $\alpha$ depends both on a fundamental physics parameter (the dimensionless coupling $\zeta$ of the scalar field to the electromagnetic sector) and on the background dark cosmology parameters. For the classes of models we have studied these are the dimensionless dark energy density $\Omega_\phi$ and the dark energy equation of state $w_0$, the latter being the crucial one.

We have obtained new, tighter constraints on the dimensionless coupling $\zeta$, and also studied their model dependence. Our results indicate that current constraints on Class I models will be dominated by the atomic clock bound of \cite{Rosenband}. However, this is not the case for Class II models---the BSBM model provides an explicit example of this point. The constraints are somewhat dependent on the choice of (flat or logarithmic) prior for the dark energy equation of state, but regardless of this choice these constraints are non-trivial and competitive.

We note that different currently available astrophysical measurements of $\alpha$---specifically the archival data of Webb {\it et al.} and the dedicated measurements of Table \ref{table1}---lead to somewhat different constraints, with the former leading to a mild (statistically not significant) preference for non-zero couplings. For the combined measurements we always find results consistent with the standard paradigm. In any case these discrepancies highlight the importance of obtaining improved astrophysical measurements of $\alpha$ (both in terms of statistical uncertainty and in terms of control over possible systematics), not only for their own sake but also because there can have a strong impact on dark energy studies. The ongoing UVES Large Program should further improve the status quo, and the next generation of high-resolution ultra-stable spectrographs such as ESPRESSO (due for commissioning in late 2016) and ELT-HIRES will be ideal for this task. A broad roadmap for these studies is outlined in \cite{cjmGRG}, and some specific forecasts of the future impact of these measurements may be found in \cite{Leite,LeiteNEW}.

In the classes of models under consideration the new degree of freedom inevitably couples to nucleons (through the $\alpha$ dependence of their masses) and leads to violations of the Weak Equivalence Principle. We have therefore used our bounds on $\zeta$ to derive indirect bounds on the E\"{o}tv\"{o}s parameter $\eta$. Despite the aforementioned model dependence, our indirect bounds are stronger than the current direct ones, in some cases by as much as one order of magnitude: in other words, they are at the $\eta\sim$few$\times10^{-14}$ level. We note that the forthcoming MICROSCOPE mission, currently scheduled for launch in April 2016, should reach $\eta\sim 10^{-15}$ sensitivity \cite{MICROSCOPE}. Should this measure a value of $\eta$ larger than that in our bounds, this would rule out the Class I models we have studied here (or alternatively would imply that the measurements of $\alpha$ on which they rely are incorrect and dominated by systematics).

Finally, we note that the forthcoming generation of high-resolution ultra-stable spectrographs will also provide significantly tighter constraints on $\eta$. Specifically, for Class I models and based on the forecasts of \cite{Leite}, we may expect a sensitivity of $\eta\sim$few$\times10^{-16}$ for ESPRESSO and $\eta\sim 10^{-18}$ for ELT-HIRES. The latter is similar to the expected sensitivity of the proposed STEP satellite \cite{STEP}. Therefore null results from STEP and the E-ELT would force any putative coupling of light scalar fields to the standard model to be unnaturally small, implying that either WEP violating fields do not exist at all or that these couplings are suppressed by some currently unknown symmetry mechanism ---whose existence would be as exciting and significant as that of a light scalar field itself.

\appendix
\section{Appendix: Including Oklo}

The Oklo natural nuclear reactor also provides a complementary probe of the stability of fundamental couplings. In particular, it nominally provides a strong constraint on $\alpha$, but it only does so if one assumes that everything else is not varying. This is very a poor assumption, as has been amply documented in the recent literature. We refer the interested reader to a recent review on the subject \cite{OKLOrev} and references therein. It is clear that this is not as 'clean' and reliable a measurement as the atomic clock and QSO measurements, and for this reason we have not used it in the main text of this work. 

Nevertheless, one may legitimately ask what impact the $\alpha$ constraint on Oklo, taken at face value, would have on our analysis. This constraint, from the analysis of \cite{OKLO1}, is
\begin{equation} \label{oklo}
\frac{\Delta\alpha}{\alpha}=(0.5\pm6.1)\times10^{-8}\,,
\end{equation}
at an effective redshift $z_{\rm Oklo}=0.14$. This nominally strong bound ultimately exploits the presence of a 97.3 meV resonance in the neutron capture by the Samarium-149 isotope (whereas the typical energy scale of nuclear reactions is of order MeV). We note that even stronger bounds have been obtained in \cite{OKLO2,OKLO3}, but these rely on additional assumptions; in what follows we will use the more conservative bound of \cite{OKLO1}.

That said we can revisit the analysis done in Sect. \ref{sec:sgz} for the model of Slepian {\it et al.}, now also including the Oklo constraint. The thawing class of models (of which this model is an example) is the one for which Oklo should have a larger relative effect, since in these models the deviations from $w=-1$ are larger at low redshifts. In freezing models (or even the model with a constant equation of state ($w(z)=w_0$) the impact of Oklo will be smaller.

%%%%%%%%%%%%%%%%%%%%%%%%%%%%%%%%%%%%%%%%%%%%%%%%%%%%%%%%%%%%%%%%%%%%%%%%%%%%%%%%%%
\begin{figure}[tbp]
\centering
\includegraphics[width=3in]{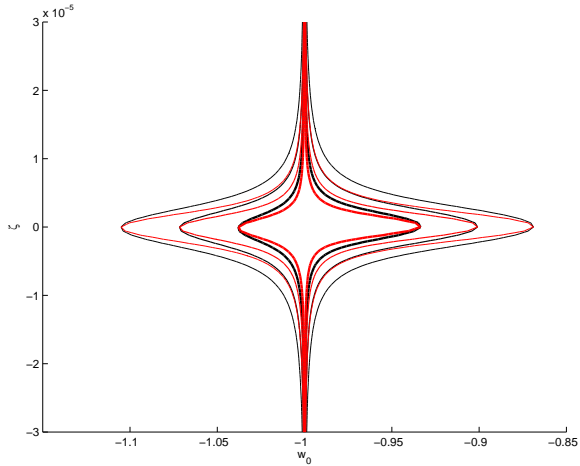}
\caption{\label{figA1}One, two and three sigma constraints on the $\zeta-w_0$ plane for the model of Slepian {\it et al.} \protect\cite{SGZ}, for the full dataset considered in the main part of our analysis (solid black lines, cf. Fig. \protect\ref{fig04}) and for the same dataset augmented by adding the Oklo bound (red dashed lines). In both cases the reduced chi-square of the best fit is $\chi^2_{min,full}=0.97$.}
\end{figure}
%%%%%%%%%%%%%%%%%%%%%%%%%%%%%%%%%%%%%%%%%%%%%%%%%%%%%%%%%%%%%%%%%%%%%%%%%%%%%%%%%%
\begin{figure}[tbp]
\centering
\includegraphics[width=3in]{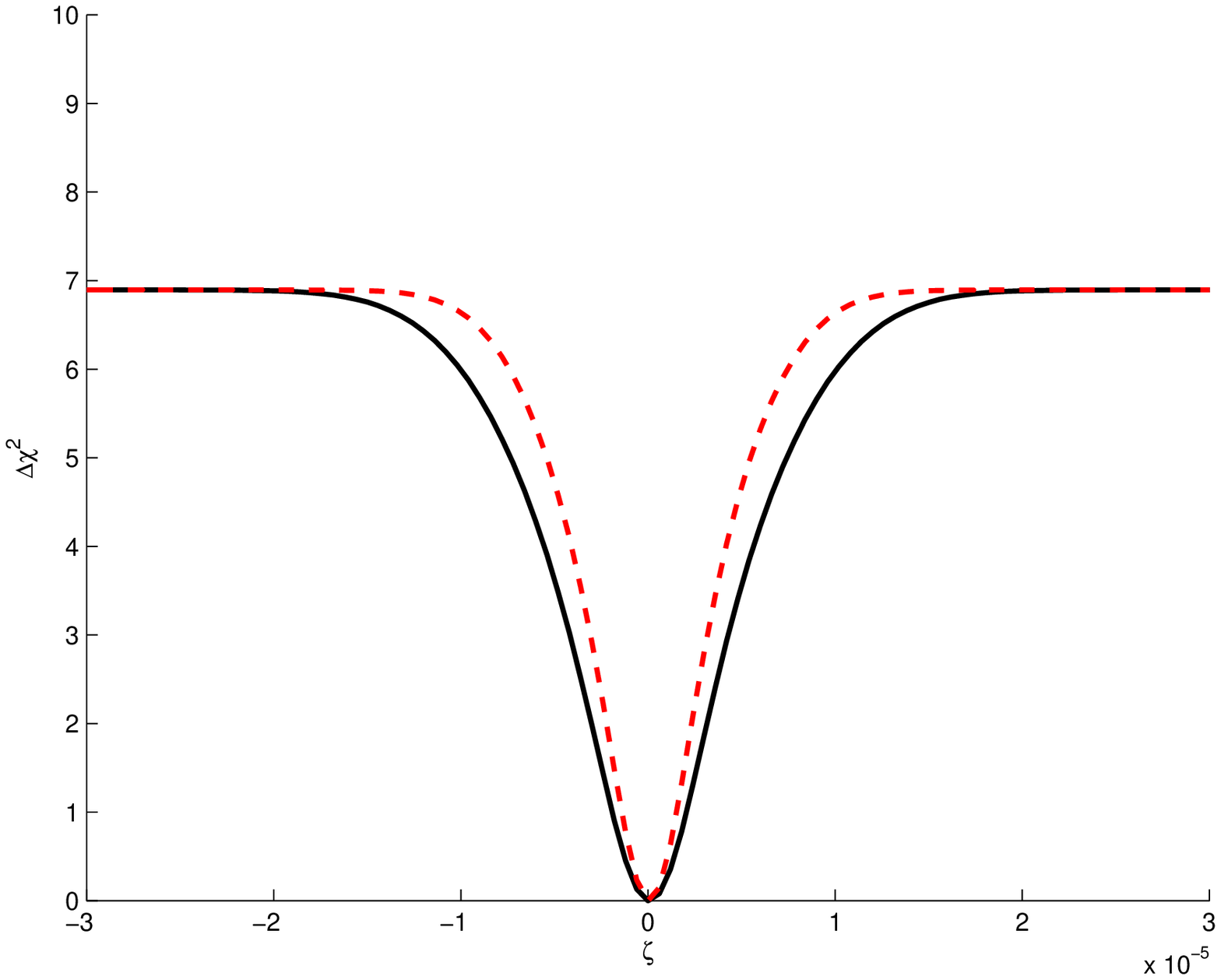}
\includegraphics[width=3in]{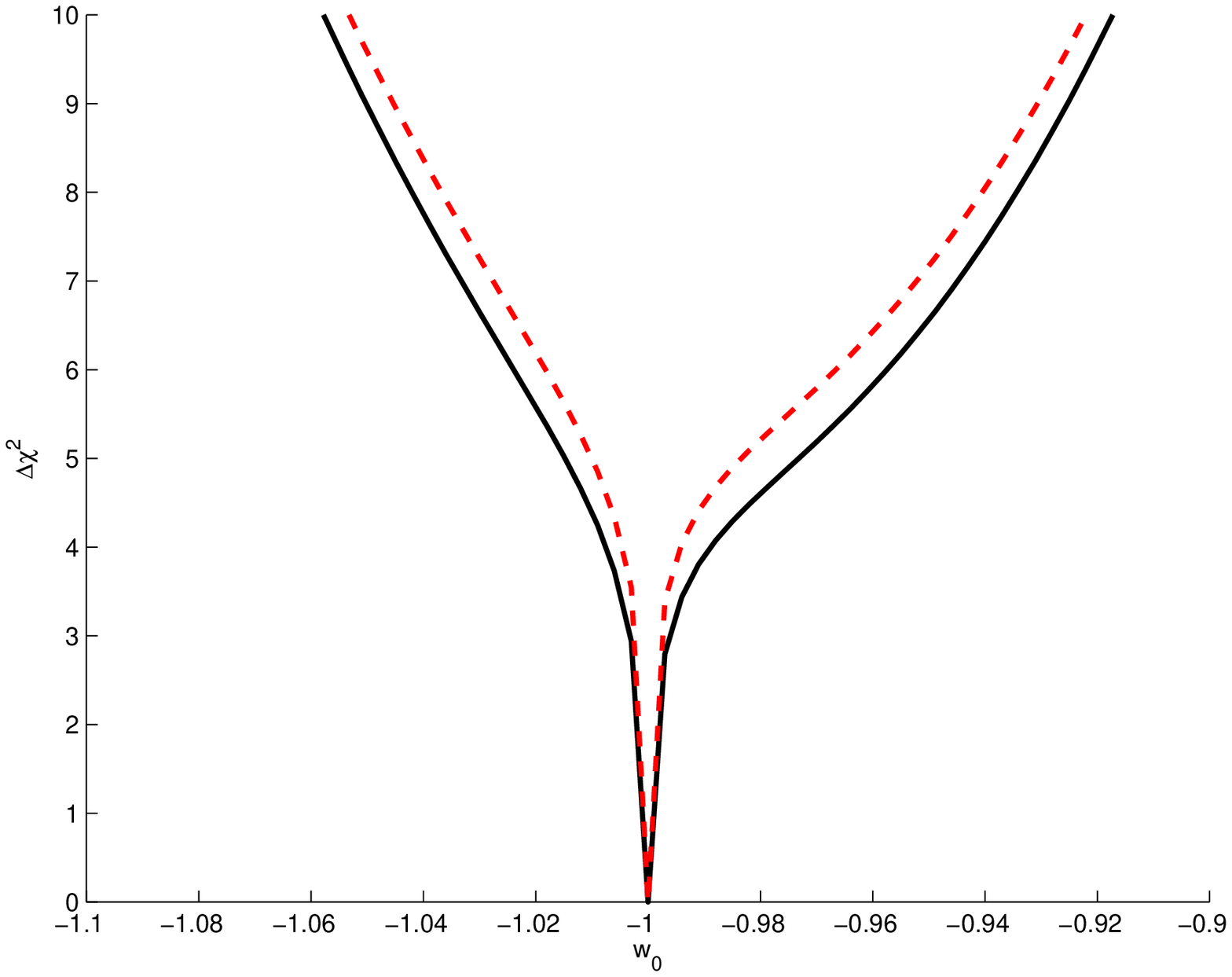}
\caption{\label{figA2}1D likelihood for $\zeta$ marginalizing over $w_0$ (left panel) and for $w_0$  marginalizing over $\zeta$ (right panel), for the model of Slepian {\it et al.} \protect\cite{SGZ}. Plotted is the value of $\Delta\chi^2=\chi^2-\chi^2_{min}$, for the full dataset considered in the main part of our analysis (solid black lines, cf. Fig. \protect\ref{fig05}) and for the same dataset augmented by adding the Oklo bound (red dashed lines).}
\end{figure}
%%%%%%%%%%%%%%%%%%%%%%%%%%%%%%%%%%%%%%%%%%%%%%%%%%%%%%%%%%%%%%%%%%%%%%%%%%%%%%%%%%

Figures \ref{figA1} and \ref{figA2} depict the Oklo impact: we show the 2D and 1D marginalized likelihoods for $\zeta$ and $w_0$, with and without adding Oklo to our datasets. Table \ref{tableA1} compares the derived constraints in the two cases. One sees that the effects of the Oklo constraint are certainly discernible, but by no means dramatic. The qualitative reason for this is that this is a constraint at very low redshift (as compared to the astrophysical measurements), and only a factor of $~3$ stronger than the atomic clock constraint. Thus a reasonable fraction of models that fit the latter constraint also fit the former.

%%%%%%%%%%%%%%%%%%%%%%%%%%%%%%%%%%%%%%%%%%%%%%%%%%%%%%%%%%%%%%%%%%%%%%%%%%%%%%%%%%
\begin{table}[tbp]
\centering
\begin{tabular}{|c|c|c|c|}
\hline
Parameter & Confidence level & Without Oklo & With Oklo \\
\hline\hline
Coupling & 95.4\%  & $|\zeta_{\rm SGZ}|<5.6\times10^{-6}$  & $|\zeta_{\rm SGZ}|<4.5\times10^{-6}$  \\
\hline
E\"{o}tv\"{o}s & 95.4\% & $\eta_{\rm SGZ}<3.1\times10^{-14}$  & $\eta_{\rm SGZ}<2.0\times10^{-14}$  \\
\hline
Eq. of State & 99.7\% & $-1.05<w_0<-0.92$ &  $-1.04<w_0<-0.93$  \\
\hline
\end{tabular}
\caption{\label{tableA1}Constraints on the relevant parameters of the model of Slepian {\it et al.} \protect\cite{SGZ}. The middle column shows the constraints discussed in Sect. \ref{sec:sgz}, which do not include the Oklo bound. The right column shows the same constraints also including the Oklo constraint of \protect\cite{OKLO1}, interpereted as a constraint on $\alpha$.}
\end{table}
%%%%%%%%%%%%%%%%%%%%%%%%%%%%%%%%%%%%%%%%%%%%%%%%%%%%%%%%%%%%%%%%%%%%%%%%%%%%%%%%%%

\acknowledgments
We are grateful to Ana Catarina Leite, David Corre and Pauline Vielzeuf for many helpful discussions on the subject of this work. This work was done in the context of project PTDC/FIS/111725/2009 (FCT, Portugal). CJM is also supported by an FCT Research Professorship, contract reference IF/00064/2012, funded by FCT/MCTES (Portugal) and POPH/FSE (EC). M.P. and M.v.W. acknowledge financial support from Programa Joves i Ci\`encia, funded by Fundaci\'o Catalunya-La Pedrera.

\end{document}